\documentclass[10pt,journal,compsoc]{IEEEtran}
\usepackage{hyperref}
\usepackage[update,prepend]{epstopdf}
\usepackage{balance}
\usepackage{amsmath}
\usepackage{caption}
\usepackage{multirow}
\usepackage{subcaption}
\usepackage{footnote}
\usepackage[linesnumbered, ruled]{algorithm2e}
\usepackage{algpseudocode}

\usepackage{amsfonts}
\usepackage{yyy-math}
\usepackage{textcase}
\usepackage{wrapfig}
\usepackage{graphicx}
\usepackage{xcolor}
\usepackage{subcaption}
\usepackage{epstopdf}
\usepackage{url}
\usepackage{tabularx}
\usepackage{paralist}
\usepackage{amssymb}
\usepackage{pifont}

\newcommand{\system}{\textsc{LocationSpark}\xspace}

\newtheorem{definition}{Definition}
\newtheorem{theorem}{Theorem}

\newtheorem{proposition}{Proposition}

\usepackage{listings}
\usepackage{color}
\definecolor{dkgreen}{rgb}{0,0.6,0}
\definecolor{gray}{rgb}{0.5,0.5,0.5}
\definecolor{mauve}{rgb}{0.58,0,0.82}
\title{ LocationSpark: In-memory Distributed Spatial Query Processing and Optimization}
\author{Mingjie~Tang,Yongyang Yu, 
        Walid~G.~Aref,~\IEEEmembership{Fellow,~IEEE,} \\
        Ahmed~R.~Mahmood, 
        Qutaibah~M.~Malluhi,
        Mourad~Ouzzani,~\IEEEmembership{Member,~IEEE,}
\IEEEcompsocitemizethanks{\IEEEcompsocthanksitem M. Tang, Y. Yu and A. Mahmood are with the Department
of Computer Science, Purdue University, IN 47906.\protect\\
E-mail: tang49@purdue.edu
\IEEEcompsocthanksitem W.G. Aref is with the Department of Computer Science and Center
for Education and Research in Information Assurance and Security (CERIAS), Purdue University.
\IEEEcompsocthanksitem Q. Malluhi is with KINDI Center for Computing Research, Qatar University, Doha, Qatar.
\IEEEcompsocthanksitem M. Ouzzani is with the Qatar Computing Research Institute, HBKU, Doha, Qatar.
}% <-this % stops a space
\thanks{}}
\begin{document}

\setlength{\abovedisplayskip}{3pt}
\setlength{\belowdisplayskip}{3pt}
\setlength{\belowcaptionskip}{-5pt}

\IEEEcompsoctitleabstractindextext{
\begin{abstract}
Due to the ubiquity of spatial data applications and the large amounts of spatial data that these applications generate and process, 
there is  a pressing need  for  scalable spatial query processing. 
In this paper, we present new techniques for spatial query processing and optimization in an in-memory and distributed setup to address scalability.
More specifically, we introduce new techniques for handling query skew, which is common in practice,
and optimize communication costs accordingly. 
We propose  a distributed query scheduler that use a new cost model 
to optimize the cost of spatial query processing. 
The scheduler generates query execution plans that minimize the effect of query skew.  
The query scheduler employs new spatial indexing techniques based on bitmap filters to forward queries to the appropriate local nodes.
Each local computation node is responsible for optimizing and selecting its best local query execution plan based on the indexes and the nature of the spatial queries in that node. 
All the proposed spatial query processing and optimization techniques are prototyped inside Spark, 
a distributed memory-based computation system. 
The experimental study is based on real datasets and demonstrates that distributed spatial query processing can be enhanced by up to an order of magnitude over existing in-memory and distributed spatial systems.
\end{abstract}

}

\maketitle

\IEEEdisplaynotcompsoctitleabstractindextext

\IEEEpeerreviewmaketitle

\section{Introduction}
\begin{sloppypar}

Spatial computing is becoming increasingly important with the proliferation of mobile devices. 
In addition, the growing scale and importance of location data have driven the development of numerous specialized
spatial data processing systems, e.g., SpatialHadoop~\cite{spatialhadoop}, Hadoop-GIS~\cite{HadoopGIS} and MD-Hbase~\cite{MDHbase}. By taking advantage of the power and cost-effectiveness of MapReduce, 
these systems typically outperform spatial extensions on top of relational database systems by 
%MO:Why this reference?
%Mingjie: because we have an assumption here and I hope to clarify this. 
orders of magnitude~\cite{HadoopGIS}.
%These 
MapReduce-based systems allow users to run spatial queries using predefined high-level spatial operators without worrying about fault tolerance or computation distribution.
However, these systems have the following two main limitations: 
(1)~they do not leverage the power of distributed memory, and (2)~they are unable to reuse intermediate data~\cite{ZahariaPHDThesis}.
Nonetheless, data reuse is very common in spatial data processing.
For example, spatial datasets, e.g., Open Street Map (OSM, for short, $>$60G) and 
Point of Interest (POI, for short,  $>$20G)~\cite{spatialhadoop}, are usually large.
It is unnecessary to read these datasets continuously from disk (e.g., using HDFS~\cite{HDFS}) to respond to user queries.
%WGA:New--- Need reference for HDFS to be placed immediately after HDFS above.
%Mingjie: 
%AR what queries? you have not talked about your queries at all, given that you give novel algorithms for gueris, i think you may need to talk about the queries you address
%Mingjie: this query is general case of query, it for any kinds of query
Moreover, intermediate query results have to be written back to HDFS, thus  directly impeding the performance of further data analysis steps. 

One way to address the above challenges
is to develop an efficient execution engine for large-scale spatial data computation based on a memory-based computation framework (in this case, Spark~\cite{ZahariaPHDThesis}). 
Spark is a computation framework that allows users to work on distributed in-memory data without worrying about data distribution or fault-tolerance. Recently, various Spark-based systems have been proposed for spatial data analysis, e.g., SpatialSpark~\cite{SpatialSpark}, GeoSpark~\cite{GeoSpark},   Magellan~\cite{Magellan}, Simba~\cite{Simba} and LocationSpark~\cite{LocationSpark}.

%introduction of locationspark

\begin{figure}[t]
        \centering
        \begin{subfigure}[b]{0.23\textwidth}
                \includegraphics[width=1.65in]{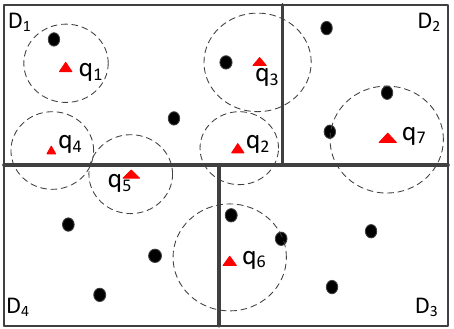}
                \caption{Spatial Range Join}
                \label{fig:sjoin}
        \end{subfigure}%
        \begin{subfigure}[b]{0.23\textwidth}
                \includegraphics[width=1.65in]{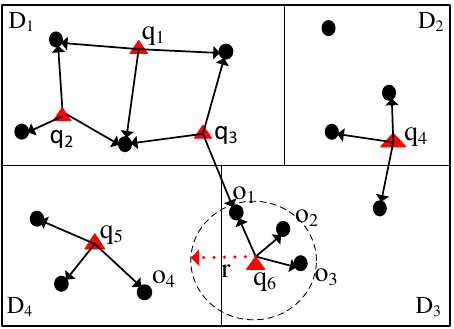}
                \caption{$k$NN Join}
                \label{fig:knnjoin}
        \end{subfigure}
        \caption{Illustration of spatial join operators. Circles centered around the triangle focal points form one dataset, and the black dots form the second dataset. (a)~Spatial range join returns the (dot,triangle) pairs when the dot is inside the circle. (b)~kNN join returns (triangle,dot) pairs when the dot is among the 3 nearest dots to the triangle.}
        \label{fig:datadistribution}
        \vspace{-1em}
\end{figure}

Although addressing 
%While tackling 
several challenges in spatial query processing, none of the existing systems
is able to overcome 
%However, none of the existing systems have a spatial query execution plan generation that can overcome 
the computation skew introduced by spatial queries. 
``Spatial query skew" is observed in distributed environments during spatial query processing
when certain data partitions  are  overloaded by spatial queries. 
%This is termed ``spatial query skew".
Traditionally, distributed spatial computing systems (e.g.,~\cite{spatialhadoop,HadoopGIS,SpatialSpark}) 
first learn  the spatial data distribution by sampling the input data. 
Afterwards, spatial data is evenly partitioned into equal-size partitions. 
For example, in Figure~\ref{fig:datadistribution}, the data points with dark dots are evenly distributed into four partitions. 
Given the partitioned data, 
consider two  spatial join operators, namely  range and $k$NN joins, to combine two datasets, say $D$ and $Q$,
with respect to a spatial relationship. 
For each point $q \in Q$, a spatial range join (Figure~\ref{fig:sjoin}) returns data points in  $D$ that are inside the radius of the circle centered at $q$. % \in Q$. 
In contrast, a $k$NN join (Figure~\ref{fig:knnjoin}) returns the $k$ nearest-neighbors from the dataset $D$ for each query point $q \in Q$. 
Both spatial operators are expensive 
%operations 
and may incur 
%expensive 
computation skew in certain workers, thus greatly degrading the overall performance. % of query processing.
%, which brings the whole query processing time degrade greatly.

For illustration, consider a large spatial dataset, with millions of points of interests (POIs), that is
%is preprocessed and 
partitioned into different computation nodes based on the spatial distribution of the data, e.g., one data partition represents data from San Francisco, CA, and another one for Chicago, IL. %, etc.
%Later on, spatial queries (e.g., range search) are , and the spatial join or $k$NN join operators combine the queries and partitioned data in parallel. 
%Mourad: This is a good attempt to give a concrete example but it's not natural! Rush hours is not correlated with looking for hotels! You need to be more creative.
%Mingjie: maybe restaurant is better motivation, since people need to hangle out
Assume that we have  incoming queries from people looking for different POIs, e.g.,  restaurants, train or bus stations, and grocery stores, 
around their locations. These spatial range queries are  consolidated into 
%Mourad: Why batches? the reader may wonder.
%Mingjie: it is impractical to query the dataset for each query, thus, we use the batch to share the execution.
%different 
batches to be joined via an index on the POI data (e.g., using indexed nested-loops join). 
%Mourad: The example is still not natural, if you live in SF you know where you're going; I don't see millions of queries being issued for restaurants in one region!
%Mourad: I tried to "diversify" the query but if someone can come up with a better example that will be great.
%Mingjie: got it
After partitioning the incoming spatial queries based on their locations, we observe the following issues:
%Mourad: Be more concrete on why this is the case; is it because there are a lot of cars on the road!
%Mourad: Mention actual queries that will show such a behavior
%Mingjie: I will user the uber data to illustrate this, thanks. 
%Mourad: Good idea!
%YYY: it seems that there is a bug here: do you mean 4pm to 6pm in Chicago, and that's early in the afternoon in SF?
%Mingjie: it is updated. thanks
During rush hours in San Francisco from 4PM to 6PM~(PST), San Francisco's data partition may encounter more queries than the data partition in Chicago, since it is already the evening in Chicago. 
Without an appropriate optimization technique, the data partition for San Francisco will take much longer time to process its corresponding queries while the workers responsible for the other partitions are lightly loaded. 
As another example, in Figure~\ref{fig:datadistribution}, the data points (the dark dots) 
correspond to Uber car's GPS records where multiple users (the triangles) are looking for the Uber carpool service around them. 
Partition $D_1$,  %corresponding 
which corresponds to an airport,   experiences more queries than other partitions because people 
%prefer Uber during 
may prefer using Uber at this location.
%Mourad: What's "traffic connection ports"?
%traffic connection ports. 
%Mourad: You're repeating yourself and it's not clear what you're trying to say here!
%Mingjie: I use the uber example here. 
%AR one issue with this example, is that is requires real-time processing, what is the average latency here, as far as i understand, you batch queries, this may impose delay, you need to say that in this case the delay is acceptable for real-time applications
%Mingjie: good idea, I will mention this later in the section 2.2.  
%Actually, 
Being aware of the spatial query skew provides a new opportunity to optimize queries
in distributed spatial data environments.
The skewed partitions 
%with the skew 
have to be assigned more computation 
%powers 
power to reduce the overall processing time.
%Being aware of the spatial query skew in distributed spatial data computation, the skewed data partitions have to be assigned more computation power than others. 
%Thus, we can reduce the processing time in query skewed partitions and improve the overall performance. 
%However, existing systems ignore such spatial query skew, and hence lose the optimization opportunity.

Furthermore, communication cost, generally a key factor of the overall performance, may become a bottleneck.
When a spatial query touches more than one data partition, 
it may be the case that some of these partitions do not contribute to the final query result. 
For example, in Figure~\ref{fig:sjoin}, queries $q_2$, $q_3$, $q_4$, and $q_5$  overlap more than one data partition ($D_1$, $D_2$, and $D_4$), but these partitions do not contain data points that satisfy the queries. 
Thus, scheduling queries (e.g., $q_4$ and $q_5$) to the overlapping data partition $D_4$ incurs unnecessary communication cost. 
More importantly, for the spatial range join or $k$NN join operators over two 
%big
large datasets, 
%the network communication cost 
the cost of network communication
may become prohibitive without proper optimization.

In this paper, we introduce~\system, an efficient memory-based distributed spatial query processing system. 
In particular, it has a query scheduler to mitigate query skew. 
The query scheduler uses a cost model to analyze the skew for use by the spatial operators, 
%AR the query scheduler does not build a cost model, it uses a cost model to represent the work load at any given node
%Mingjie: updated
and a plan generation algorithm to construct a load-balanced query execution plan. % for a query processing job. 
After plan generation, local computation nodes select the proper algorithms to improve their local performance based on the available spatial indexes and the registered queries on each node.
Finally, to reduce the communication cost when dispatching queries to their overlapping data partitions,
\system adopts a new spatial bitmap filter, termed sFilter, that can 
%detect whether a spatial object is inside the spatial range or not.
speed up query processing by avoiding needless communication with data partitions that do not contribute to the query answer. 
We implement \system  as a library in Spark that provides an API for spatial query processing and optimization based on Spark's standard dataflow operators. 
%\system requires no modifications to Spark, revealing a general method to combine spatial data processing within distributed dataflow frameworks.

The main contributions of 
%our work 
this paper  are as follows:
\begin{enumerate}
\item We develop a new spatial computing system for efficient  processing of spatial queries in a distributed in-memory environment.

\item We address  data and query skew issues to improve load balancing while executing spatial operators, 
e.g., spatial range joins and $k$NN joins, by generating cost-optimized query execution plans over in-memory distributed spatial data.
%Mourad: give few details
%Mingjie: updated

\item We introduce a new lightweight yet efficient spatial bitmap filter to reduce communication cost. %, and show its benefit in distributed spatial computation.

\item We realize the introduced query processing and optimization techniques inside Spark. We use the developed prototype system, \system, to conduct a large-scale evaluation on real spatial data and 
common benchmark algorithms, and compare~\system against state-of-the-art distributed spatial data processing systems. Experimental results illustrate an enhancement in performance by up to an order of magnitude over existing in-memory  distributed spatial systems.
%WGA: added some summary of the results. Same as in abstract. 

\end{enumerate}

The rest of this paper proceeds as follows.
Section~\ref{section-preliminaries} presents the problem definition and an overview of distributed spatial query processing. Section~\ref{section:scheduler} introduces 
%the algorithm to generate the optimal execution plan.
the cost model and the cost-based query plan scheduler and optimizer and their corresponding algorithms. 
Section~\ref{section:locatlPlan} presents an empirical study for local execution plans in local computation nodes. Section~\ref{section:sftiler} introduces the spatial bitmap filter, and explains how it can speedup spatial query processing in a distributed setup. The experimental results are presented in Section~\ref{section:evaluation}. Section~\ref{section:relatework} 
%introduces 
discusses
the related work. Finally, Section~\ref{section:conclusion} concludes the paper.
\end{sloppypar}

\section{Preliminaries}
\label{section-preliminaries}
\subsection{Data Model and Spatial Operators}
\system stores spatial data as key-value pairs. A 
tuple, say $o_i$, contains a spatial geometric key $k_i$ and a related value $v_i$.  
The spatial data type for  key $k_i$ can be a two-dimensional point, e.g., latitude-longitude, a line-segment, a poly-line, a rectangle, or a polygon. 
The value type $v_i$ is specified by the user, e.g., a text data type if the data tuple is a tweet. 
In this paper, we assume that queries are issued  
%MO: What do you mean by progressively?
%Mingjie: continuously. 
continuously by users, and are processed by the system in 
%MO: Explain batches?
batches (i.e., similar to the DStream model~\cite{ZahariaPHDThesis}).

\system supports 
%a rich set 
various types 
of spatial query predicates including spatial range search, $k$-NN search, spatial range join, and $k$NN join. 
%Moreover, it supports data updates and spatio-textual operations. 
In this paper, we focus our discussions on the spatial range join and $k$NN join operators
on two dataset, say $Q$ and $D$, which form the outer and inner tables, respectively. %, of the  join operators.  
%These two operators are prohibitively expensive especially when processing big spatial data. 

\sloppypar
\begin{definition}\textbf{Spatial Range Search} -  $range(q,D)$:
Given a spatial range area $q$ (e.g., 
%radius
circle or rectangle) and a dataset $D$, 
$range(q,D)$ finds all tuples from $D$ that overlap the spatial range defined by $q$. 
\end{definition}

\sloppypar
\begin{definition}\textbf{Spatial Range Join} - $Q \Join_{\text{sj}}D$:
 Given two dataset $Q$ and $D$, 
 $Q \Join_{\text{sj}}D$, combines each object $q \in Q$ with its range search results from $D$,  $Q \Join_{\text{sj}}D$= $\{ (q,o) | q \in Q, o \in range(q,D)\}$.
\end{definition}

%Mourad: Check the usage of \forall in your definitions? 

%Yongyang: Move sj and knn to the subscript of the join operators.
\begin{definition}\textbf{$k$NN Search} - $k$NN(q,D):
Given a query tuple $q$, a dataset $D$,  and an integer $k$, 
%the $k$ nearest neighbor search,  
$k$NN(q,D), returns the output set $\{o | o \in D$ and 
$\forall s \in D$ and $ s \neq o, ||o, q|| \leq ||s, q||\}$, where the number of output objects from $D$, $|kNN(q, D)|$ = $k$.
\end{definition}

\begin{definition}\textbf{$k$NN Join} - $Q \Join_{\text{knn}}D$:
Given a parameter $k$, $k$NN join of $Q$ and $D$ computes each object $q \in Q$ with its $k$ nearest neighbors from $D$. 
$Q \Join_{\text{knn}}D$= $\{ (q,o) | \forall q \in Q, \forall o \in kNN(q,D)\}$.

\end{definition}

\begin{figure}[ht]
   \centering
\includegraphics[width=2.5in, height=1.5in]{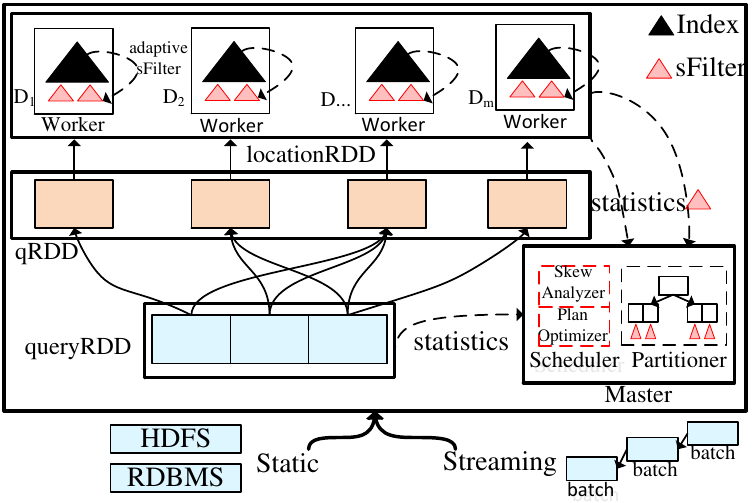}
\caption{\system system architecture}
\label{fig:system_overview}
\end{figure}

%\begin{figure}[ht]
 %  \centering
%\includegraphics[width=3in]{pics/datadistribution}
%\caption{Spatial operators and data distribution}
%\label{fig:datadistribution}
%\end{figure}

\subsection{Overview of In-memory Distributed Spatial Query Processing}
\label{section:systemoverview}

\begin{sloppypar}

%AR illustrates the architecture of~\system
%Mingjie:updated
%If there is no index on the data in the distributed memory,
%AR i suggest remove Naturally
%Mingjie: updated
%spatial queries have to go through all data tuples.
%AR I would write it like this, "if there is no index, we would have to iterate over all data tuples" however this is trivial 
%Mingjie: updated
%Mourad:The process of building the index is not clear at all!
%Mingjie: the build index is a very standard approach for build index over distributed data. (1) sample data to get the data distributions. (2) system partition input data into N partitions, and each partitions owns the similar amount of data (3) then, each workers build local index for the partitioned data. I follow the some well known written to illustrate this procedure. 
%YYY: what do you mean N leaves sequentially? And each worker processes its own piece of initial data partition of D. Here N is also the number of workers, right?
%Mingjie: the master node build the index with N leaves with centralized approach
To 
%enhance 
facilitate 
spatial query processing, we build a distributed spatial index for in-memory spatial data.
%via the following steps.
Given 
%an input 
%MO: The reader may be lost on how come an index built using a sample is good to find data?
a spatial dataset $D$, we obtain samples from $D$ and construct a spatial index (e.g., an R-tree)
%Quadtree)
with $N$ leaves over the samples.
%with $N$ leaves, sequentially. 
We refer to this index on the sample data as
the {\em global spatial index}.
%Mourad: The data is partitioned once so it cannot be partitioned into N partitions by each worker!
%Mingjie: the data is read from some other sources, then each worker get parts of data, and then worker redistribute data based on its location into $N$ parts. 
Next, each worker  partitions the %input 
dataset $D$ into $N$ partitions according to the built global spatial index via data shuffling. 
%AR  what is data shuffling
%Mingjie: this is very basic in distributed computation, shuffling means propagate something.
The global spatial index guarantees that each data partition approximately has the same amount of data.
%AR this statement depends on the index type, a bad index may result in imbalance 
%Mingjie: the build index is learned from sample data, and is a standard way for data partitioning. 
%YYY: each executor takes D/N amount data, right?
%Mingjie: yes, 
Then, each worker $W_i$ of the $N$ workers has a local data partition $D_i$ that is roughly $1/N$th of the data and builds a local spatial index.
%AR this statement is not clear, and why  $1/N$
%Mingjie: since we have N partitions, and each partition take 1/N data
%AR this is not clear, you may say it as a percentage or something 
Finally, the indexed data (termed the  LocationRDD) is cached into memory. 
Figure~\ref{fig:system_overview} gives the architecture of~\system and the physical representation of the partitioned spatial data based on the procedure outlined above, 
where the master node stores the global spatial index that indexes the data partitions, while each worker has a local spatial index over the local spatial data within the partition. 
Notice that the global spatial index  partitions the data into LocationRDDs as in Figure~\ref{fig:system_overview}, and this index can be copied into various workers to help partition the data in parallel. 
%Mourad: The following sentence is not clear.
%Mingjie: its updated.
%Mourad: Vague but ok.
%MO: It will be good is you can give a concrete example of a specific application and how you select the type of the index.
%Mingjie: updated
The type of each local index,  e.g., a Grid, an R-tree, or an IR-tree, for a data partition can be determined based on the specifics of the application scenarios. 
For example, a Grid can be used for moving objects indexing while an R-tree can be used for polygons objects.  
%In order to support data update, 

%AR this is the first time you define an executor, you need to define it first 
%Mingjie: executor is like the worker. 

%AR add citations here to r-tree and quad-tree
%AR there should be a discussion on how you choose the index, do you let the administrator specific the type of index used, or you choose the index autamtically?? since you are using multiple indexes, there should be a discussion on the guidlines on the index usage
%Mingjie:  this is putted in the section 4. 

%Mourad: Be more clear on what's happening, replication is to move Query points and data points to the workers; the way you wrote it is confusing.
%Mourad: You need to explain the setting master node, workers, where data is stored, how it is replicated, etc. and then the process.
%Mingjie: it is updated. 
%Given a set of spatial queries, say $Q$, the query scheduler is responsible for creating a query execution plan for $Q$ over the distributed indexed data. 
For  spatial range join, two strategies are possible; either replicate the 
%queries of 
outer table and send it to the node where the inner table is or replicate the inner table data and send it to the different processing nodes where the outer table tuples are.
%queries are.
In a shared execution, the outer table is typically a collection of range query tuples and the inner table is the queried dataset. 
If this is the case, it would make more sense to send the outer table of queries to the inner data tables 
as the outer table is usually much smaller compared to the inner data tables.
In this paper, we adopt the first approach because it would be impracticable to replicate and forward copies of the large inner data table. 

Thus, each tuple $q \in Q$ is replicated and forwarded to the partitions that spatially overlap with it. 
These overlapping partitions are 
identified using the global index. 
Then, a post-processing step merges the local results to produce the final output. 
%YYY: do you mean q_5 or q_2?
%Mingjie: updated
For example, we replicate the outer table $q_2$ in Figure~\ref{fig:sjoin}  and forwarded it to data partitions $D_1$, $D_3$, and $D_4$.
Then, we execute a spatial range search on each data partition locally. 
Next, we merge the local results to form the overall output of tuple $q$. 
%Mourad: Why "queries" and not "query", what do you mean by "re"partitioned, It's simply the same  query that is replicated on different partitions?
%Mourad: It should be simpler and also try to use more intuitive naming conventions 
%Mingjie: it is updated, the repartiioned based on the queries and data partitions overlapping. this is not replicated. 
As illustrated in Figure~\ref{fig:system_overview}, 
the outer table that corresponds to a shared execution plan's collection of queries (termed queryRDD) are first partitioned into 
%MO: What's qRDD?
qRDD based on the overlap between the queries in qRDD and the corresponding data partitions. 
Then, local search takes place over the local data partitions of LocationRDD.
%WGA: Mingjie, I see it unacceptable to argue about replicating the data and moving it to the queries. You are claiming that this is for big Spatial data. So, saying that you will replicate and forward the data sounds unacceptable. Let me know what you think.
%WGA: I omitted the second strategy as it does not make sense.
%Mingjie: yes, the second approach is used by others, and it is impractical to replicate data for each spatial join. 
%In the second strategy,  in contrast to replicating the queries, we replicate and forward the data to the queries. In particular, data is  replicated and is forwarded into a query partition if the data is within a distance $r'$ 
%to 
%from
%the boundary of that partition, 
%Mourad: How do you get $r$?
%Mingjie:　the way to compute is explained in the where condition
%Mourad: You mean circle not rectangle, a rectangle doesn't have a radius!
%Mourad: Also, why are you talking about "queries" and not "query"?
%Mingjie: the queries means that there are multiple query for a  join execution. 
%where $r'$ is the maximum radius of a query circle that make up the queries. 
%In this paper, we adopt the first approach since it is impracticable to replicate large amounts of data for different incoming queries. 

%AR you did not mention in the perlimanties section, that you are having batches of queries and data. i belive that data is incremental and queries arrive in batches.
%Mingjie: yes, the queries are arrive in batches. and the data update is also support, but this work mainly focus on incoming queries. 

The $k$NN join operator is implemented similarly in a simple two-round process.
%with small modifications. 
%Mourad: shuffled you mean moved?
%Mingjie: yes, in spark, we can move data via data shuffling. 
%AR the problem is we know this term, as it is used in map-reduce systems, but some others reviewer may not be aware of it
%Mingjie: no, shuffle data is a general term in distributed system. 
First, each outer focal points $q_i \in Q$ is transferred to the worker that holds the data partition that $q_i$ spatially belongs to. 
Then, the $k$NN join is executed locally in each data partition,  producing the $k$NN candidates for each focal point $q_i$. 
Afterwards, the maximum distance from $q_i$ to its $k$NN candidates, say radius $r_i$, is computed. 
If the radius $r_i$  overlaps 
%with 
multiple data partitions, point $q_i$ is replicated into these overlapping partitions, and another set of $k$NN candidates is computed in each of these partitions.
Finally, we merge the $k$NN candidates from the various partitions to get the exact result. 
For example, in Figure~\ref{fig:knnjoin}, assume that we want to evaluate a $3$NN query  for Point $q_6$.
The first step is to find the $3$NN candidates for $q_6$ in data Partition $D_3$.
Next, we find that the radius $r$ for the $3$NN candidates from Partition $D_3$ overlaps Partition $D_4$.
Thus, we need to compute the $3$NN of $q_6$ in Partition $D_4$ as well. 
Notice that the radius $r$ can enhance the $3$NN search in Partition $D_4$ because only the data points within Radius $r$ are among the $3$NN of $q_6$. 
Finally, the $3$NN of $q_6$ are $o_1$, $o_2$ and $o_3$. 
\end{sloppypar}
%\begin{figure}[ht]
%   \centering
%\includegraphics[width=2.7in]{pics/framework}
%\caption{LocationSpark is a layer on top of Spark (lines of code)}
%\label{Fig:System}
%\end{figure}

\subsection{Challenges}
\label{section-challenges}

\begin{sloppypar}
%Mourad: The observation you're making here is not a consequence of the previous paragraph/example. It refers back to the example mentioned in the intro. You need to have sequence in your ideas.
%Overall, we observe that even if the input data $D$ is evenly distributed in different workers, certain workers still suffer from more spatial queries than others. 
%This always happens 
%Mingjie: yes, it is, but I use several small sentence to illustrate our ideas. 
The outer and inner tables (or, in shared execution terminology, the queries and the data) are spatially collocated in distributed spatial computing. In the following discussion, we  refer to the outer table as being the queries table, e.g., containing the ranges of range operations, or the focal points of $k$NN operations. 
We further assume that the outer  (or queries) table is the smaller of the two.
We refer to the inner table  by the data table 
(in the case of shared execution of multiple queries together).
The distribution of the incoming spatial  
queries (in  the outer tables) changes dynamically over time, with bursts in certain spatial regions. Thus, evenly distributing the input data $D$ to the various workers may result in load imbalance at times. 
%as some  workers will have more spatial queries than others at different times. 
\system's scheduler 
identifies the skewed data partitions 
%Mourad: How?
%Mingjie: based on the cost estimation. 
based on a cost model and then repartitions and redistributes the data accordingly, and selects the optimal  repartitioning strategies for both the outer and inner tables, and consequently generates an overall optimized execution plan. 

Communication cost is a major factor that affects system performance. 
\system 
%propose to use  
adopts 
%an efficient 
a spatial bitmap filter  to reduce network communication cost.
The spatial bitmap filter's role is to prune the data partitions that overlap %with 
the spatial ranges from the outer tables but do not contribute to the final operation's results. 
This spatial bitmap filter 
%has to be 
%is persistent in memory and 
is memory-based and
%needs to be 
is space- and time-efficient. 
The spatial bitmap filter %also needs  to adapt
adapts its structure %w.r.t 
as the data and query distributions change.

%WGA: I am not sure that we need the paragraph below. Let me know if it is needed.
%Overall, the goal of \system~is to support spatial data processing efficiently at very large scale while achieving performance speedup over existing systems such as SpatialSpark, GeoSpark and Magellan, by optimizing several components.
\end{sloppypar}

%MO: Try to use dotted lines instead of red
\begin{figure}[ht]
   \centering
\includegraphics[width=3in, height=1.3in]{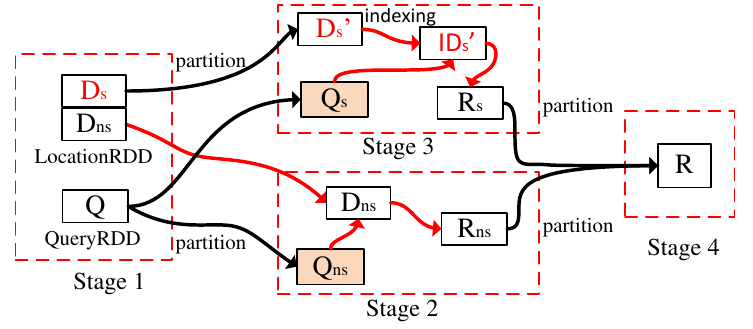}
\caption{Execution plan for spatial range join. The red lines identify  local operations, and black lines show the data partitioning. $D_s$ and $D_{ns}$ are the skew and non-skew  partitions, respectively. Queries $Q$ (the outer table) are partitioned into skew $Q_s$ and non-skew $Q_{ns}$ in Stage 1. Stages 2 and 3 execute  independently. Stage 4 merges the results.}  
\label{fig:queryplan}
\vspace{-1.8em}
\end{figure}

\section{Query Plan Scheduler}
\label{section:scheduler}

This section addresses how to dynamically handle query (outer table) skew.
First, we present the cost functions for query processing and analyze the bottlenecks. 
Then, we show how to repartition the skewed data partitions to speedup processing. This is formulated as an optimization problem 
that we show is NP-complete. 
Consequently, we introduce an approximation algorithm to solve the skew data repartitioning problem. 
Although presented for spatial range joins, the proposed technique  applies to $k$NN join as well. 
%, since (1) repartition technique here can mitigate the query skew for computing $k$NN candidates in each partition, (2) computing the $k$NN candidate from data partition, which are overlapped by radius $r$, is similar as spatial join.

%the skew analyzer and optimized plan generation are designed for general spatial queries processing.   
%AR i think you may need to say why you need a cost function here?, i mean that the cost function is important to resresent the workload at any given worker, and you aim to achieve balance accross workers 
%AR you mention that your cost functions focus on the join operators, what about range select and knn select ? i think you may need to have a general representation of the cost function of a partition, and that this cost function changes according to registered queiries, 
%Mingjie: yes, the cost function here is general type of cost function, and you can find that I mention the cost estimation is depend on the queries types, index and other factors later. 
%AR aslo, do you allow the execution of several query types at the same batch, i.e., KNN join with spatial join or not, because, if yes then if you use an index that works for one operation, it may not work for the other 
%Mingjie: for query type in the same batch should be the same. one index can work for knn and spatial join over indexed data. 

\subsection{Cost Model}
%MO: Try to stay consistent, you already decided that inner is data table and outer is query table
%MO: Also, either table or dataset
The input table $D$ (i.e., inner table of spatial range join) is distributed into $N$ data partitions, and each data partition $D_i$ is indexed and cached in memory. 
For the query table $Q$ (i.e., outer table of spatial range join), each query $q_i \in Q$ is shuffled to the data partitions that spatially overlap with it. 
The shuffling cost is denoted by $\epsilon(Q, N)$. 
The execution time of local queries at Partition $D_i$ is $\E(D_i)$. 
%is $\E(D_i)=\E(|D_i|, |Q_i|)$, where $|D_i|$ and $|Q_i|$ are the number of data (inner) points and (outer) queries at Partition $D_i$. 
The execution times of local queries depend on the queries and built indexes, and the estimation of $\E(D_i)$ is presented later.
After the local results are computed, the post-processing step merges these local results to produce the final output. The corresponding cost is denoted by $\rho(Q)$.

Overall, the runtime cost for the spatial range join operation is:
\begin{equation} \label{eq:costfunction}
C(D,Q)=\epsilon(Q, N)+ \max_{i \in [ 1, N]} \E (D_i)+ \rho(Q),
\end{equation}
where $N$ is the number of data partitions. In reality, the cost of query shuffling is far less than the other costs as the number of queries is much smaller than the number of data items.
Thus, the runtime cost can be estimated as follows:
\begin{equation} \label{eq:appcostfunction}
C(D,Q)=\max_{i \in [ 1, N]} \E (D_i)+ \rho(Q)
\end{equation}
%In Equation~\ref{eq:appcostfunction}, 
We categorize the data partitions into two types:  skewed  ($D^{s}$) and non-skewed ($D^{ns}$). The execution time of the local queries in the  skewed partitions is the bottleneck.
The runtime costs for skewed and non-skewed data partitions are $ \max_{i \in [ 1, \hat{N} ]}  \E(D_i^{s}) $ and $ \max_{j \in [ 1, \bar{N} ]}   \E(D_j^{ns})$, respectively, where $\hat{N}$ (and $\bar{N}$) is the number of skewed (and non-skewed) data partitions, and $N=\hat{N}+ \bar{N}$. Thus, Equation~\ref{eq:appcostfunction} can be rewritten as follows:
\begin{equation} \label{eq:skewcostfunction}
C(D,Q) =\max \{ \max_{i \in [ 1, \hat{N} ]}  \E(D_i^{s}) , \max_{j \in [ 1, \bar{N} ]}  \E(D_j^{ns}) \}+ \rho(Q)
\end{equation}

\subsection{Execution Plan Generation}

The goal of the query scheduler is to minimize the query processing time subject 
to the following constraints: 
(1)~the limited number of available resources (i.e., the number of partitions) in a cluster, and
(2)~the overhead of network bandwidth and disk I/O. 
Given the partitioned and indexed spatial data,  
the cost estimator for query processing based on sampling that we  introduce below, 
and the available number of data partitions, 
the optimizer returns an execution plan that  minimizes query processing time.
First, the optimizer  determines if any partitions are skewed.
Then, it repartitions them subject to the introduced cluster and networking constraints. 
Finally, the optimizer evaluates the plan on the newly repartitioned data to determine whether it minimizes query execution time or not (Refer to  Figure~\ref{fig:queryplan}). 

%At first, the 
Estimating the runtime cost of executing the
%for query processing (e.g., 
local queries and the cost of merging the final results is not straightforward.
%hard to give a precise number. 
The local query processing time $\E(D_i)$ is influenced by various factors including
the types of spatial indexes used, 
the number of data points in $D_i$, the number of queries directed to $D_i$, 
related spatial regions,
%WGA: What do you mean by: related spatial regions?
%Mingjie: the overlapped data partitions. 
%WGA: This is still unclear to me. Is it the number of something? State what it refers to. Given D_i, there are no spatial regions that overlap with it other than the spatial regions of the queries. Am I correct? In this case, do you mean the number of such queries, or their total area of coverage that overlap the partition or what?
%Mingjie: you are correct
and the available memory. 
Similar to~\cite{SkewHandler}, we  assume that the related cost functions  are monotonic,
and can be approximated using  samples from 
the outer and inner tables (the queries and the datasets tables). 
Thus, the local query execution time is formulated as follows:
$\E (D_i)= \E_s(\tilde{D_i}, \tilde{Q_i},  \alpha, A)$, 
%\Rightarrow R$, 
where 
$\tilde{D_i}$ is a sample of the original inner table dataset, 
$\tilde{Q_i}$ is the sample of the outer table queries,  
$A$ is the area of the underlying spatial region, 
and $\alpha$ is the sample ratio to scale up the estimate to the entire dataset.  
After computing a representative sample of the data points and queries, 
%MO: did you ever study the effect of using different sampling methods?
%Mingjie: actually, I just use the basic sampling approach provided by spark
e.g., using reservoir sampling~\cite{SamplingVitter85}, 
the cost function $\E(D_i)$ estimates the query processing runtime in data partition $D_i$. 
More details on computing  $\E (D_i)$, $\rho(Q_i)$, and the sample size can be learned from previous work~\cite{SkewHandler}.
%Mourad: This entire paragraph seems useless now! As we don't know how to compute E(D_i)! You keep repeating the same thing without telling us how to sample and then computer this estimate.
%
%AR do you have a discussion on the sample size?
%Mingjie: this is not focus on this work, the sample size is a factor to influence the estimation accurancy. 
%AR as far as i understand in spark, some data is in memory and some in desk, the repatitioning cost of memory data is less than the that of those on desk.. do you account for this fact?
%Mingjie: this is good question. actually, we assume that the cost estimation do not consider the storage level of data, and it  gives the approximate number based on sampling. 
\begin{sloppypar}

%YYY: check the following formula, use D_i or D_i hat? need to be consistent, especially for beta and gamma.
%Mingjie Updated. 
Given the estimated runtime cost over skewed and non-skewed partitions, the optimizer splits one skewed data Partition $D_i^{s}$ into $m'$  data sub-partitions.  
Assume that $\hat{Q_i}$ is the set of queries originally assigned to Partition $D_i^{s}$.
Let the overheads due to data shuffling, re-indexing, and merging be $\beta(D_i^{s})$, $\gamma (D_i^{s})$ and $\rho(\hat{Q_i})$, respectively.
Thus, after splitting a skewed Partition $D_i^{s}$, the new runtime is:
\begin{equation} \label{eq:oneskewcostfunction}
\widehat{\E(D_i^{s})}=\beta(D_i^{s}) + \max_{s \in [1, m']} \{\gamma(D_s)+ \E(D_s)\} + \rho(\hat{Q_i}).
\end{equation}
Hence, we can split one skewed Partition $D_i^{s}$ into multiple partitions only if $\widehat{\E(D_i^{s})}$ $< \E(D_i^{s})$. As a result, the new 
query execution time, 
say 
$\widehat{C(D,Q)}$, is:
\begin{equation}\label{eq:opTcostfunction}
\widehat{C(D,Q)} =\max \{\max_{i \in [ 1, \hat{N} ]}  \{\widehat{\E(D_i^{s})} \},\max_{j \in [ 1, \bar{N} ]} \{ \E ({D_j^{ns}}) \} \}+ {\rho(\bar{Q})}. 
\end{equation}
Thus, we can formulate the query plan generation based on the data repartitioning problem as follows:
\end{sloppypar}

\begin{sloppypar}
\begin{definition}
Let $D$ be the set of spatially indexed data partitions,
$Q$ be the set of spatial queries, 
$M$ be the total number of data partitions, and
their corresponding cost estimation functions, 
i.e., 
%query processing 
local query execution $\E (D_i)$, data repartitioning $\beta(D_i)$, and data indexing cost estimates $\gamma (Q_i)$.  
The query optimization problem is to 
choose a skewed Partition $D^{s}$ from $D$, 
repartition each $D_i^{s} \in D^{s}$ into multiple partitions, 
and assign spatial queries to the new data partitions. 
The new data partition set, say $D'$, contains partitions $D_1', D_2', \ldots, D_k'$. s.t. 
(1) $\widehat{C(D,Q)} < C(D,Q)$  
and (2)  $|D'| \le M$.
\end{definition}

Unfortunately, this problem is NP-complete. In the next section, we present 
a greedy
algorithm for this problem.
\end{sloppypar}

\begin{theorem}
Optimized query plan generation with data repartitioning for distributed indexed spatial data is NP-complete.
\end{theorem}

%WGA: Too wordy.
%The details of the proof is presented in the technical report~\cite{LocationSparkTechReport}.
%MO: Why not give the proof here or at least a sketch of it.
%Mingjie: it is not easy to give a brief introduction with a few sentences
The proof is given in~\cite{LocationSparkTechReport}.

%YYY: Perhaps it's better to say we propose a greedy algorithm. Usually, for an approx. algo. we need to show ratio bound, etc.
%Mourad: Agree!
\subsection{A Greedy Algorithm}
%introduce our greedy approach and it can shows better performance

%We now present the proposed approximation algorithm. 
The general idea is to search for
%AR we are searching. This sentence is not clear, searching for what 
skewed partitions based on their local query execution time. 
%We then 
Then, we
split the selected data partitions only if the runtime can be improved. 
If the runtime cannot be improved, or if 
all
the available data partitions are 
%used up,
consumed,
%AR what do you mean by used off
%Mingjie: updated
%the procedure stops. 
the algorithm terminates.
While this greedy algorithm cannot  
%AR can not guarantee having the minimum query processing time 
%YYY: shall we mention the improvement here or in the experiment section?
guarantee optimal query performance, our experiments show significant  improvement (by one order of magnitude)
over the plan executing on the original partitions.
Algorithm~\ref{alg:GreedyPartition} gives the pseudocode for the greedy partitioning procedure.
%1. greedy chose partition to repartition,
%2. evaluation and estimating the cost for re-partitioning and run time for the related partition.
%3. add the candiates into plan, queries splitting
%4. finish the whole process, once the constraint is met

\begin{sloppypar}
%introduce the subroutine function
Algorithm~\ref{alg:GreedyPartition} includes two 
%subroutines, 
functions, namely \textit{numberOfPartitions} and \textit{repartition}. 
Function \textit{numberOfPartitions} computes the number of partitions $m'$ for splitting one skewed partition. 
Naively, we could split a skewed partition into two partitions each time. But this is not necessarily efficient. 
Given data partitions $D=\{ D_1, D_2, \ldots, D_N \}$, 
let Partition $D_1$ be
the one with the largest local query execution time $\E(D_1)$. 
From Equation~\ref{eq:appcostfunction}, the  execution time is approximated by $\E(D_1)+ \rho(Q)$. 
To improve this execution time, Partition $D_1$ is split into $m'$ partitions, 
and the query execution time for Partition $D_1$ is updated to  $\widehat{\E (D_1)}$. 
For all other partitions $D_i \in D$ ($i \neq 1$), the runtime is the $\max \{ \E (D_i) \}+ \rho(Q')=\Delta$, 
where $i =[2,\ldots, N]$ and $Q'$ are the queries related to all data partitions except $D_1$. 
Thus, the runtime is $\max \{ \Delta, \widehat{\E (D_1)}\}$, and is improved if
\begin{equation} \label{eq:numberpartition}
 \max \{ \Delta, \widehat{\E (D_1)} \} < \E(D_1)+\rho(Q)
 \end{equation}
As a result, we need to compute the minimum value of $m'$ to satisfy Equation~\ref{eq:numberpartition}, since $\Delta$, $\E(D_1)$, and $\rho(Q)$ are known.

Function \textit{repartition} splits the skewed data partitions and 
reassigns the
%the related 
spatial queries to the new data partitions using two strategies. 
The first strategy  repartitions based on the data distribution. 
Because each data partition $D_i$ is already indexed by a spatial index, the data distribution can be learned directly by recording the number of data points in each  branch of the index. 
Then, we repartition data points in $D_i$ into multiple parts based on the recorded numbers while guaranteeing that each newly generated sub-partition contains an equal amount of data. 
In the second strategy, we  repartition a skewed Partition $D_i$ based on the distribution of the  spatial queries. 
First, we collect a sample
%AR sample
%Mingjie: updated
$Q_s$ from the queries $Q_i$ that are originally assigned to partition $D_i$. Then, we compute how $Q_s$ is distributed in Partition $D_i$ by recording the frequencies of the queries as they belong to branches of the index over the data. 
Thus, we can repartition the indexed data based on the query frequencies. Although the data sizes may not be equal, the execution workload will be balanced. 
In our experiments, we choose this second approach to overcome query skew. 
To illustrate how the proposed query-plan optimization algorithm works, consider the following example.
%AR why do you mention an approach you do not use?
%Mingjie: people will argue this. 
\end{sloppypar}

\begin{sloppypar}
%YYY: Can we elaborate how to set the parameters? and explain a little bit about the splitting procedure for D1.
%Mingjie: this parameter is learned from sample data
%\begin{example}
%MO: It will be good is you can draw the example.
\textbf{Running Example.} 
Given data partitions $D= \{D_1, D_2, D_3, D_4, D_5 \}$, 
where the number of data points in each  partition is 50, 
the number of queries in each partition $D_i$, $1 \leq i \leq 5$ is 30, 20, 10, 10, and 10, respectively, 
and the available data partitions $M$ is 5.
For simplicity, the local query processing cost is $\E(D_i)$ $=$ $|D_i| \times |Q_i|  \times p_e$, 
where $p_e=0.2$ is a constant.
%scalar variable. 
The cost of merging the results
%results merging cost 
is $\rho(Q)=|Q| \times \lambda  \times p_m$, 
where $p_m = 0.05$, 
%is a scalar variable and equals 0.005, 
and $\lambda=10$ is the approximate number of retrieved data points 
%for each 
per query. 
The cost of data shuffling and re-indexing after repartitioning 
%cost and data 
is 
%MO: These two are the cost of data shuffling and re-indexing?
%Mingjie: it is updated
$\beta(D_i, m')= |D_i| \times m' \times p_r$, and $\gamma(D_s)= |D_s| \times p_x$, 
%MO:respecetively for what?
respectively, where $p_r= 0.01$ and $p_x=0.02$. Without any optimization, from Equation~\ref{eq:appcostfunction}, the estimated runtime cost for this input table is 340. 
\system optimizes the query as follows. 
At first, it chooses data Partition $D_1$ as the skewed partition to be repartitioned because $D_1$ has the highest local runtime (300), while the second largest cost is $D_2$'s (200). 
Using Equation~\ref{eq:numberpartition}, we split $D_1$ into two partitions, i.e., $m'=2$.
Thus, Function \texttt{repartition} splits $D_1$ into the two partitions $D_1'$ and $D_2'$ based on the distribution of queries within $D_1$. 
The number of data points in $D_1'$ and $D_2'$ is 22 and 28, respectively, 
and the number of queries are 12 and 18, respectively. 
Therefore, the new runtime is 
%improved
reduced to $\approx 200+25$ because  $D_1$'s runtime is reduced to $\approx 100$  based on Equation~\ref{eq:oneskewcostfunction}.
Therefore, the two new data partitions $D_1'$ and $D_2'$ are introduced in place of $D_1$. 
Next, Partition $D_2$ is chosen to be split into two partitions, and the optimized runtime becomes $\approx 100+15$. 
%WGA: Isn't it the case that one D_1 is split into D'_1 and D'_2, D_1 disappears and hence we are only consuming one additional partition? In this case, the sentence below would be wrong.
%Mingjie: the D_1 do not disappear since it still take the worker to split this partition. thus, the available partition number would be minus 2 rather 1. notice the process to run over skew and non-skewed partition at the same time,
Finally, the procedure terminates as only one available partition is left.
%AR maybe a figure would help
%\end{example}
\end{sloppypar}

\begin{algorithm}[t]\small

\KwIn{
$D$: Indexed spatial data partitions,
$\text{Stat}$: Collected statistics, e.g., the number of data points and queries in data partition $D_i$,  
$M$: number of available data partitions.
%learned cost function $\E(D_i)$, $\beta(D_i)$, $\gamma (D_i),\rho(Q)$
}
\KwOut{$Plan$: Optimized data and query partition plan, $C$: estimated query cost}

 $h$: Maximum Heap; \\
 inserts $D_i$ into $heap$ // data partitions are ordered by cost $\E(D_i)$ that is computed using $\text{Stat}$ \\
 $Cost_o$ $\leftarrow$ $\E( h.top)+\rho(Q)$ // old execution plan runtime cost \\
 $Plan$ \\
\While{$M>0$}
 {
 Var $D_x$ $\leftarrow$ $h$.pop(); //get the partition with maximum runtime cost\\
 Var $m'$ $\leftarrow$ \textit{numberOfPartitions}($h$,$D_x$, $M$)  \\
 Var ($D_s$, $PL_s$) $\leftarrow$ \textit{repartition}($D_x$, $m'$) //split $D_x$ into $m'$ partitions\\
 $Cost_x \leftarrow \beta(D_x) + \max_{s \in [1, m']} \{\gamma(D_s)+ \E(D_s)\} \}+ \rho(Q_x)$ //updated runtime cost over selected skewed partition\\

  \If{$Cost_x < Cost_o$ }
  {
	  save Partitions $D_s$ into $h$ \\
      save Partition plan $PL_s$ into $Plan$ \\
      $Cost_o$ $\leftarrow$ $Cost_x$ \\
      $M$ $\leftarrow$ $M$-$m'$ \\
  }\Else
  {
     break;
  }
 }
 \caption{\small{Greedy Partitioning Algorithm}}
%\caption{\small{Optimization Query Schedule Plan}}
%WGA: Can you rephrase the above caption? I am not sure I understand this title: Optimization Query Schedule Plan?
%Mingjie: can we have the title like "Greedy Partitioning Algorithm for Optimization Query Schedule Plan"
\label{alg:GreedyPartition} 
\end{algorithm}

%MO: Since there is no much contribution in this section, I suggest to put inside the experiment section
\section{Local Execution}
%on Computing Node}
\label{section:locatlPlan}

\begin{sloppypar}
%\textbf{Optimized local execution plan}
Once the query plan is generated, each computation node chooses a specific local execution plan based on the queries assigned to it and the indexes it has. 
We implement  various centralized algorithms for spatial range join and $k$NN join operators within each worker and study their performance. The algorithms are implemented in Spark.
%in-memory environment.
%Mourad: What do you mean by using the experimental results!? 
%Mingjie: the performance of different approaches
%The results are guided to select an optimal local execution plan. 
We use the execution time as the performance measure.
%, while the number of queries and data points in each worker are identical.
\end{sloppypar}

\subsection{Spatial Range Join}
%Spatial join is extensively studied. Sowell et.al\cite{SpatialjoinSurevy} conduct an experimental study for in-memory spatial join algorithms, and show their performance for query execution and data updating. In this work, we mainly focus on query execution.
%AR i suggest to clearly mnetion what is the purpose of this section, for example, In this section, we contrast the performance of different spatial join algorithms.
We implement two algorithms for spatial range join~\cite{SpatialjoinSurevy}. 
The first is indexed nested-loops join, where we probe the spatial index repeatedly for each outer tuple (or range query in the case of shared execution). 
The tested algorithms  are nestRtree, nestGrid and nestQtree,
%AR where  R-tree, Grid and Quadtree indexes are used respectively. 
where they use an R-tree, a Grid, and a Quadtree as index for the inner table,  respectively. 
The second algorithm 
%for spatial range join 
is based on the dual-tree traversal~\cite{DualTree}. It builds two spatial indexes (e.g., an R-tree) over both the input queries 
%MO:Avoid repeating this as we already said that IQ=OT and D=IT
%Mingjie: updated
 and the data, and performs a depth-first search over the dual trees simultaneously.
%AR i think you need citations here for those technique? other than the survey 
%AR why are listing a comparsion between those techniques while they are compared in the survey? is there a difference in perfromaance from the survey?
%Mingjie: the comparison here is designed for in-memory distributed computation system, the previous work is only on single machine. We can cite all of related work, but it seems like unnecessary. 

%Figure~\ref{fig:localSpatialJoin} shows the performance of different strategies as increasing the number of queries and data points in each worker.

\begin{figure}
        \centering
        \begin{subfigure}[b]{0.19\textwidth}
                \includegraphics[width=\columnwidth]{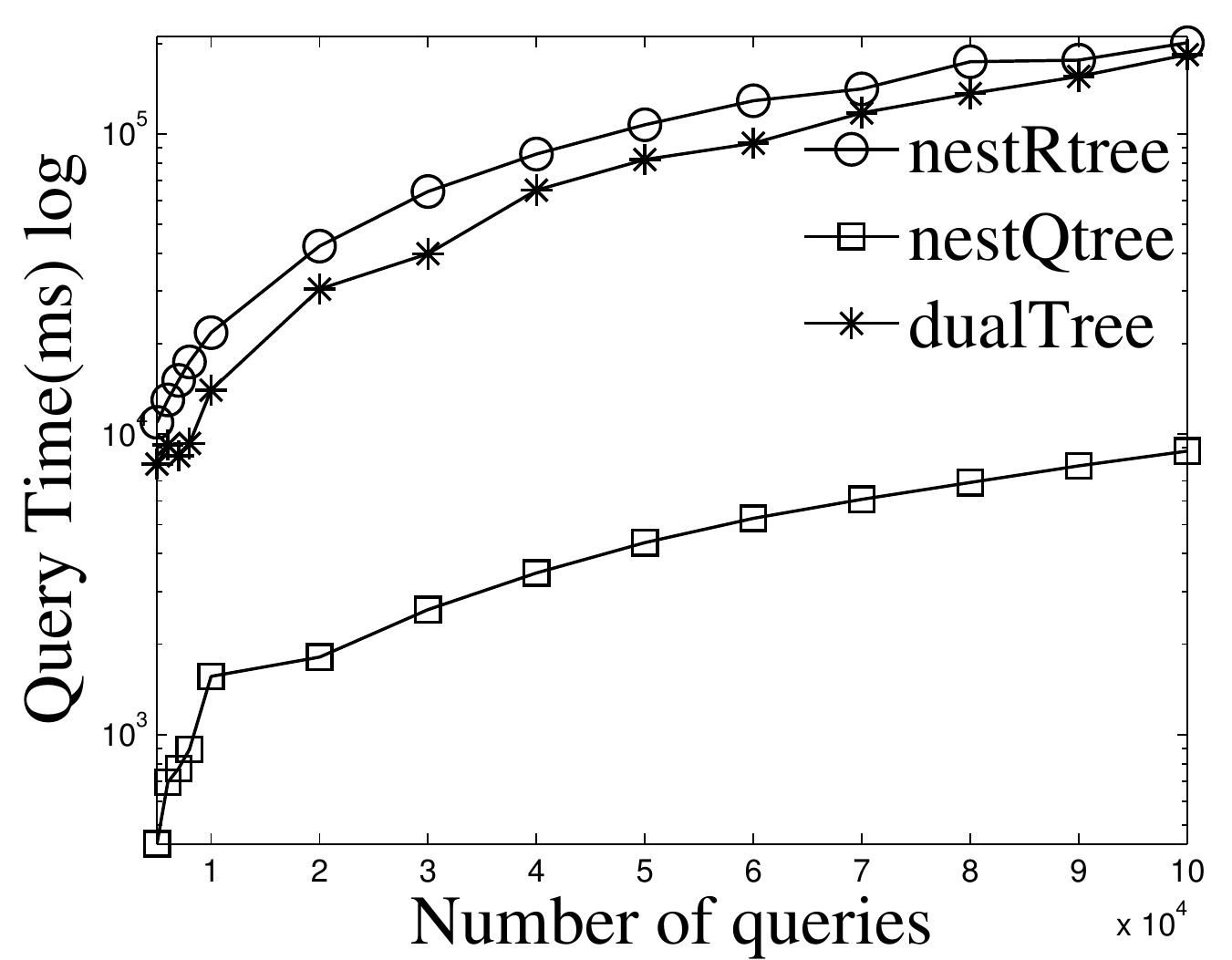}
                \caption{Effect of query size}
                \label{fig:localspatialjoin_query}
        \end{subfigure}%
        \begin{subfigure}[b]{0.19\textwidth}
                \includegraphics[width=\columnwidth]{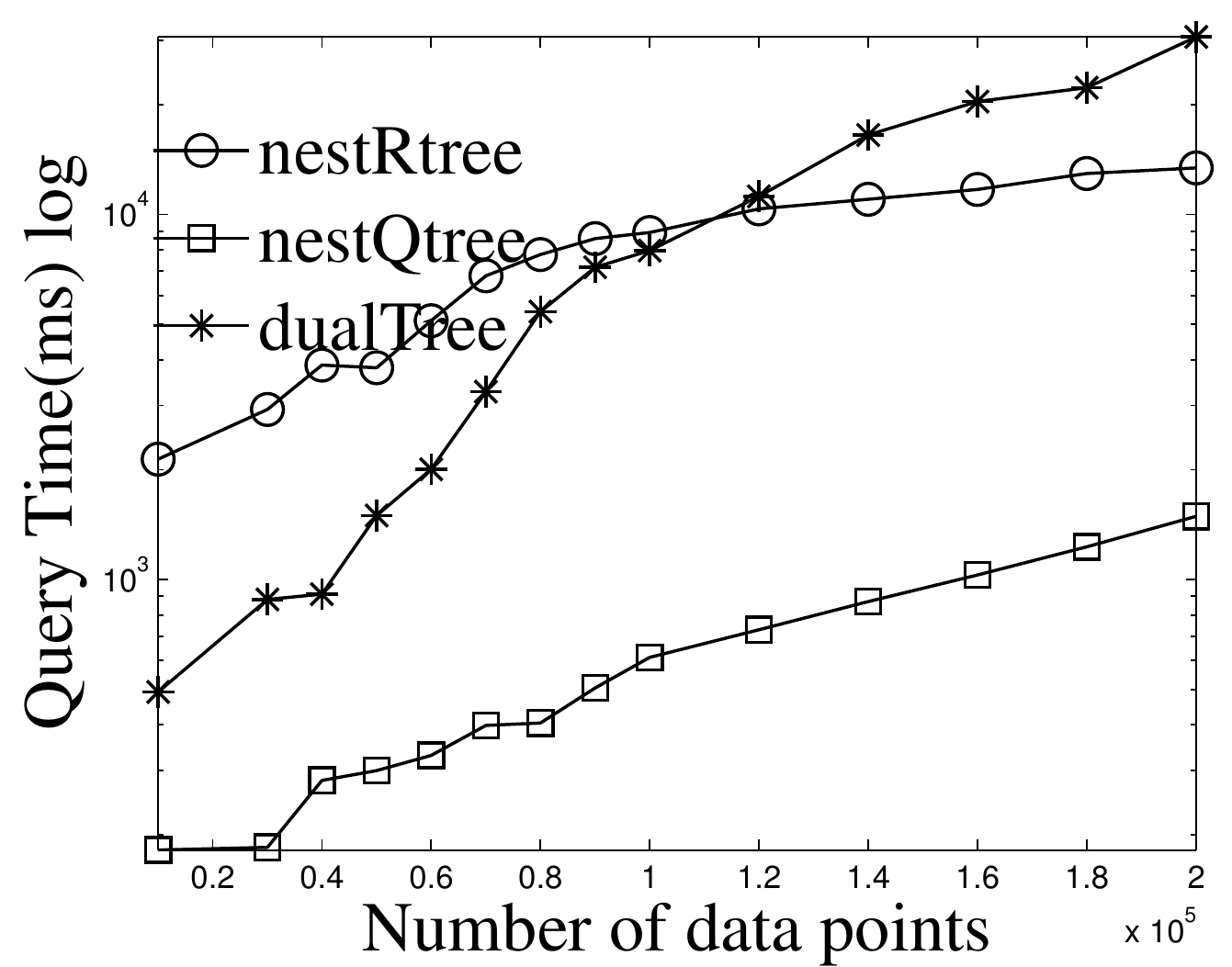}
                \caption{Effect of data size}
                \label{fig:localspatialjoin_datasize}
        \end{subfigure}
        \caption{Evaluation of local spatial join algorithms}
        \label{fig:localSpatialJoin}
        %\vspace{-1em}
\end{figure}

%MO: What's the point of putting these experiments in this section and not in the evaluation section?
Figure~\ref{fig:localspatialjoin_query} gives the performance  of 
nestRtree, nestQtree, and dual-tree, where the number of data points in each worker is 300K.
The results for nestGrid are not given as it performs the worst. 
%WGA: This is in contrast to what Ahmed Refaat shows that Grid performs always the best.
%Mingjie: yes, the grid suffer extensive skew for unbalanced data distribution.
%AR : First of all, i do not address the join query between two point datasets, Second, these  results depend on two factors.(1) what is the granularity of the grid if the grid is of sufficiently high granulation it will achieve good performance. (2) what is the range of the query. if you have very low granularity and a small spatial distance of the join query, then the performance may be bad.
%AR i recommend that when you mention the performance of the grid, you mention its granularity since some other papers advocate for using the grid, i believe you already send me a citation to one such paper. you may meet a grid advocate that may refute some of these claims  
The dual-tree approach provides a 1.8x
speedup over the nestRtree. This conforms with  other published results~\cite{SpatialjoinSurevy}.
%AR this conforms with the results reported by Sowell et. all
%AR Prof aref told to me that i should not use a citation as a part of a sentence 
nestQtree achieves an order of magnitude improvement over the dual-tree approach.
%AR achieves an order of magnitude improvement 
The reason is that the minimum bounding rectangles (MBRs)  of the spatial queries  overlap with multiple MBRs in the data index, and this reduces the pruning power of the underlying R-tree. 
%AR this reduces the pruning power of the underlying R-tree
The same trend is observed in Figure~\ref{fig:localspatialjoin_datasize} when increasing the number of indexed
%AR the number of indexed
data points. The dual-tree approach outperforms nestRtree, when the number of data points is smaller than 120k. However, dual-tree slows down afterwards. In this experiment, we only show the results for indexing over two dimensional data points. However, Quadtree performs worst when the indexed data are polygons~\cite{SametBook}. Overall, for multidimensional points, the local planner chooses nestQtree as the default approach. For complex geometric types, the local planner uses the dual-tree approach based on an R-tree implementation. 
%AR based on the these results, we adopt the nestQtree as the default spatial algorithm. However, for complex geometric types , e.g., poly-lines, the local planner uses the dual-tree algorithm.

%AR one issue here, is that you do you not show a novelty interms of the used algorithm. and you are just using one. I think you may say that we use change those centralized algorithms to work in the distributed spark platform. or something like that 

%Notice, \system also provides the option for users to specific indexes based on requirements.
\subsection{kNN Join}
\label{section:locatlknnjoinplan}

Similar to the spatial range join, indexed nested-loops can be applied to $k$NN join, where it computes the set of $k$NN objects for each query point in the outer table. An index is built on the inner table (the data table). 
The other kinds of $k$NN join algorithms are  block-based.  
They partition the queries (the outer table) and the data points (the inner table) into different blocks, and find the $k$NN candidates for queries in the same block. Then, a post-processing refine step computes $k$NN for each query point in the same block. Gorder~\cite{Gorder} divides query and data points into different rectangles based on the G-order, and utilizes two distance bounds to reduce the processing of unnecessary blocks. For example, the min-distance bound is the minimum distance between the rectangles of the query points and the data points. The max-distance bound is the maximum distance from the queries to their $k$NN sets. If the max-distance is smaller than the min-distance bound, the related data block is pruned.
%AR this G-orderd part is not very clear 
%The first distance bound is min-distance bound, it is the minimum distance between rectangle of queries and data points in the same block. The second distance bound is max-distance bound, it is the maximum distance from query points to its $k$NN sets. Naturally, if the max-distance bound is smaller than min-distance bound, the related data bock is pruned safely. 
PGBJ~\cite{PGBJ} has a similar idea that  extends to parallel $k$NN join using MapReduce. 
%PGBJ at first partitions the query and data points into different groups based on selected pivots, and the min and max-distance bound is computed based on the pivots of groups.  
Recently, Spitfire~\cite{tkdeAllkNN} 
%WGA: What is the name for this technique? Is it Spitfire or Spifilter? You use both.
%Mingjie: it would be spitfire, sorry about this. 
is a parallel $k$NN self-join algorithm for in-memory data. It replicates the possible $k$NN candidates into its neighboring data blocks. Both PGBJ and Spitfire are designed for parallel $k$NN join, but they are not directly applicable to indexed data. The reason is that PGBJ partitions queries and data points based on the selected pivots while Spitfire is specifically optimized for $k$NN self-join. 

\system enhances the performance of the local $k$NN join procedure. For the Gorder~\cite{Gorder}, instead of using the expensive principal component analysis (PCA) in Gorder, we apply the Hilbert curve to partition the query points. We term the modified Gorder method {\em sfcurve}. We modify PGBJ as follows. First, we compute the pivots of the query points based on a clustering algorithm (e.g., k-means) over sample data,
%AR sampled data 
and then partition the query points into different blocks based on the computed pivots.
%AR the computed pivots
Next, we compute the MBR of each block.
%AR Next, the rectangles of every block are computed
Because the data points are already indexed (e.g., using an R-tree), 
%AR e.g., using an R-tree
the min-distance from the MBRs of the query points and the index data is computed, and the max-distance bound is calculated based on the $k$NN results from the pivots. This approach is termed {\em pgbjk}. In terms of spitfire, we use a spatial index  to speedup finding the $k$NN candidates.

%AR i belive the following is a good contribution from your side, i think you need to highlight it more i think you may give it a name to your changes 

%AR is the sfcurve is your name fore your modification of the  Gorder algorithm ? if yes i think you should highlight this more as contribution 

\begin{figure}
        \centering
        \begin{subfigure}[b]{0.19\textwidth}
                \includegraphics[width=\columnwidth]{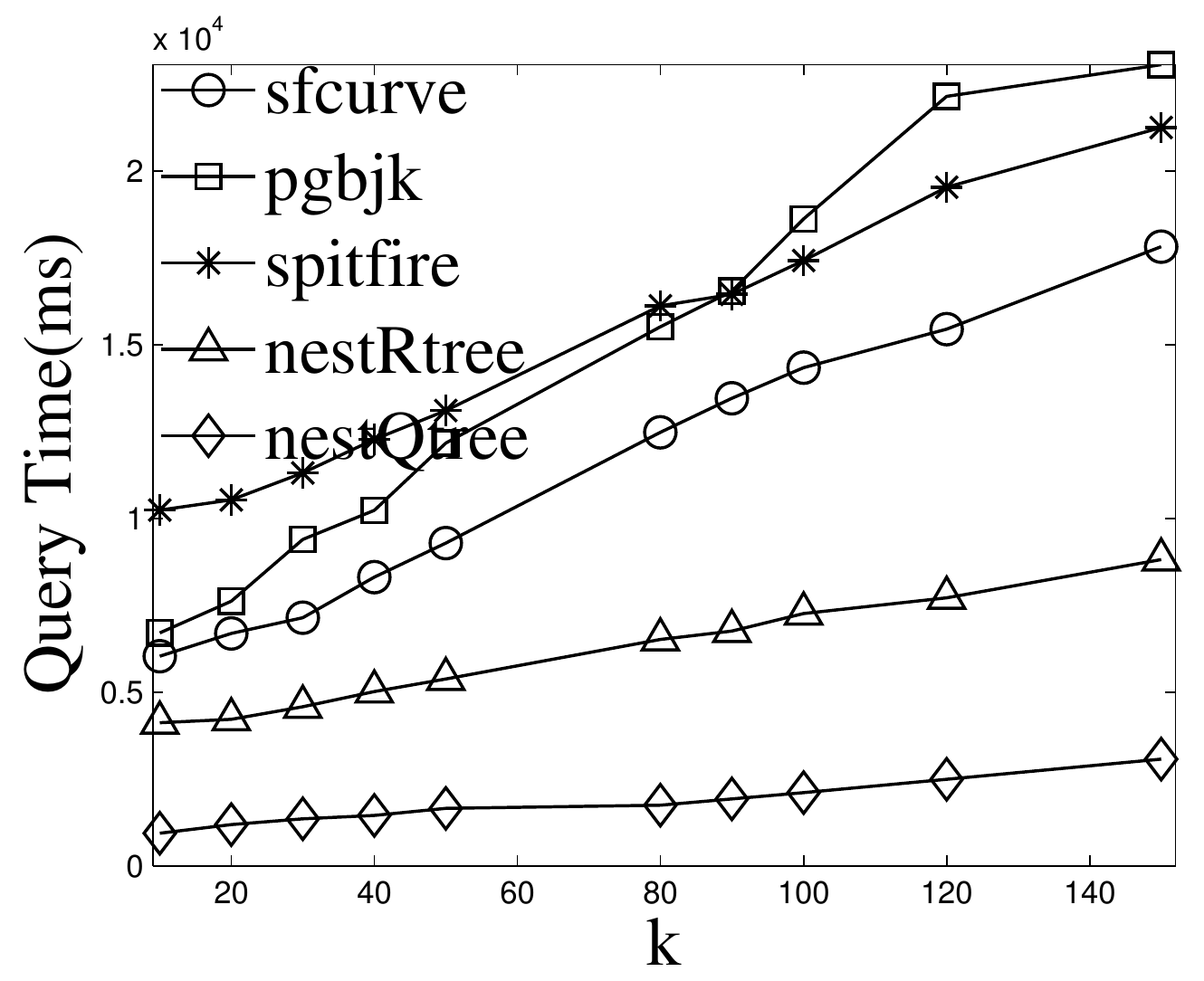}
                \caption{Effect of $k$}
                \label{fig:localknnjoin_k}
        \end{subfigure}%
        \begin{subfigure}[b]{0.19\textwidth}
                \includegraphics[width=\columnwidth]{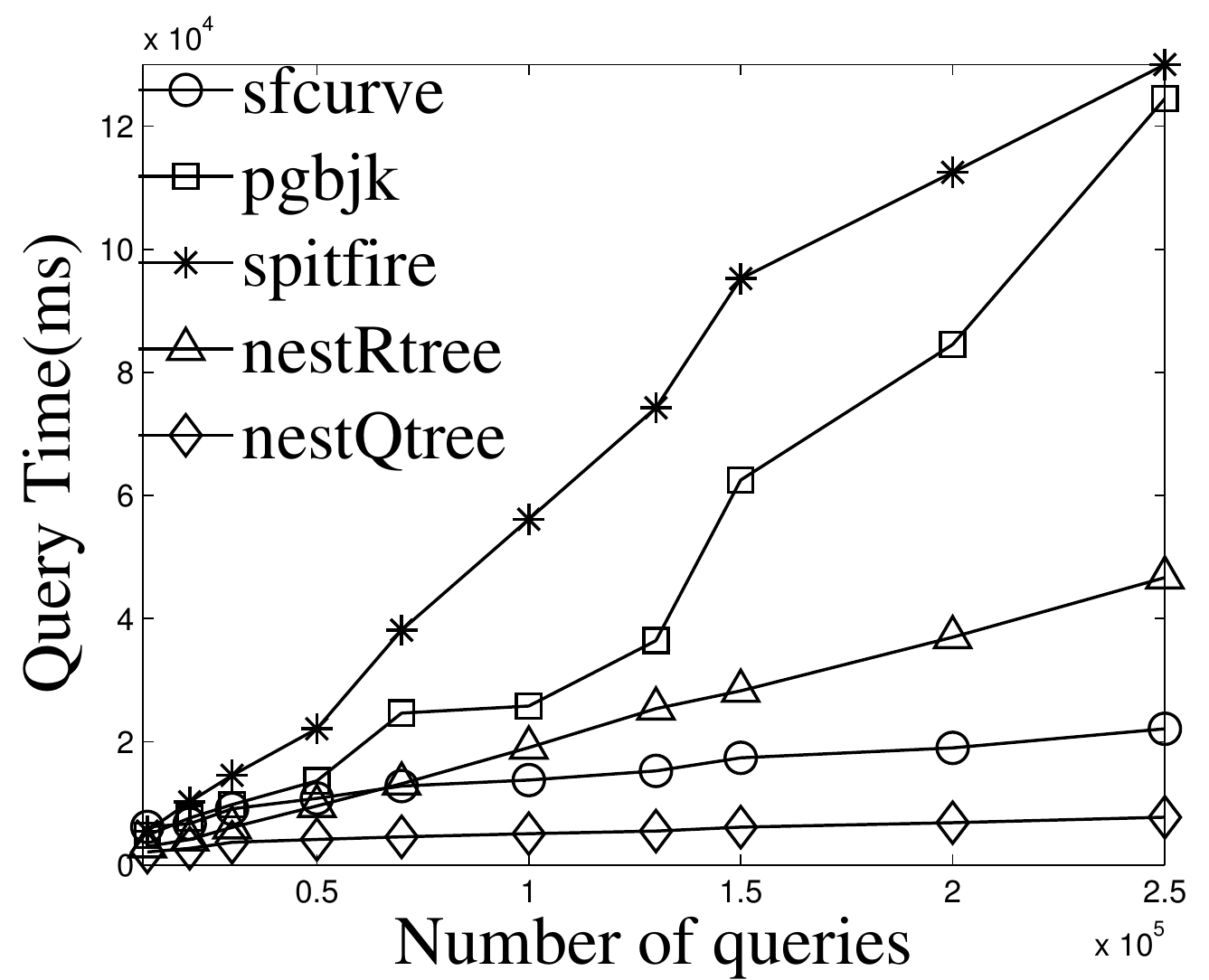}
                \caption{Effect of query size}
                \label{fig:localknnjoin_datasize}
        \end{subfigure}
        \caption{Evaluation of local $k$NN join algorithms}
        \label{fig:localKnnJoin}
        \vspace{-1em}
\end{figure}

Figure~\ref{fig:localknnjoin_k} gives the performance of the specialized $k$NN join approaches within a local computation node when varying $k$ from 10 to 150. The nestQtree approach always performs the best, followed by nestRtree, sfcurve, pgbjk, and spitfire. Notice that block-based approaches induce extensive amounts of $k$NN candidates for query points in the same block, and it directly degrades the performance of the $k$NN refine step. More importantly, the min-distance bound between the MBR %rectangle 
%WGA: Mingjie, is use of MBR above correct in place of rectangle?
%Mingjie: yes, it is correct. 
of the query points and the MBR of the data points is much smaller than the max-distance boundary. Then, most of the data blocks cannot be pruned, and result in redundant computations. In contrast, the nested-loops join algorithms prune some data blocks because the min-distance bound from the query point to the data points of the same block is bigger than the max-distance boundary of this query point. Figure~\ref{fig:localknnjoin_datasize} gives the performance of the $k$NN join algorithms by varying the number of query points. Initially, for less than 70k query points, nestRtree outperforms sfcurve, then nestRtree degrades linearly with more query points. Overall, we adopt nestQtree as the $k$NN join algorithm for local workers.

\section{Spatial bitmap filter}
\label{section:sftiler}
%Bloom filter is widely used for testing whether a data tuple is contained in a set or not.
In this section, we introduce a new spatial bitmap filter termed  {\em sFilter}.  
%The sFilter can answer whether a specific spatial data point that lies inside a spatial range actually exists in the underlying data set or not. 
The sFilter helps us  decide for an outer tuple, say $q$, of a spatial range join, if there exist tuples in the inner table that actually join with $q$. This helps reduce the communication overhead. For example, consider an outer tuple $q$ of a spatial range join
%(a.k.a. spatial join)
where $q$ has a  range that overlaps multiple data partitions of the inner table. Typically, all the overlapping partitions need to be examined by communicating $q$'s range to them, and searching the data within each partition to test for overlap with $q$'s range. This incurs high communication and search costs.
Using the sFilter, given $q$'s range that overlaps multiple partitions of the inner table, the sFilter can decide which overlapping partitions contain data that overlaps $q$'s range without actually communicating with and searching the data in the partitions. Only the partitions that contain data that overlap with $q$'s range are the ones that will be contacted and searched. 

%In the remaining of this section, we introduce the data structures for the sFilter and its corresponding range search algorithm. Next, we describe how sFilter reduces unnecessary communication costs as the scheduler assigns a query to local computation nodes. Finally, we present an approach to update the sFilter adaptively when the distribution of data and queries change over time.
%AR You may rewrite it as follows; Changes in data and query workload may result in a degradation of the  sFilter performance, i.e., an increase in the number of false positives in query results. To this end, we propose an adaptive sFilter update algorithm that is sensitive to changes in workload.

\subsection{Overview of the sFilter}
\label{section:sFilterOverview}

\begin{figure}[ht!]
   \centering
\includegraphics[width=3.3in, height=1.8in]{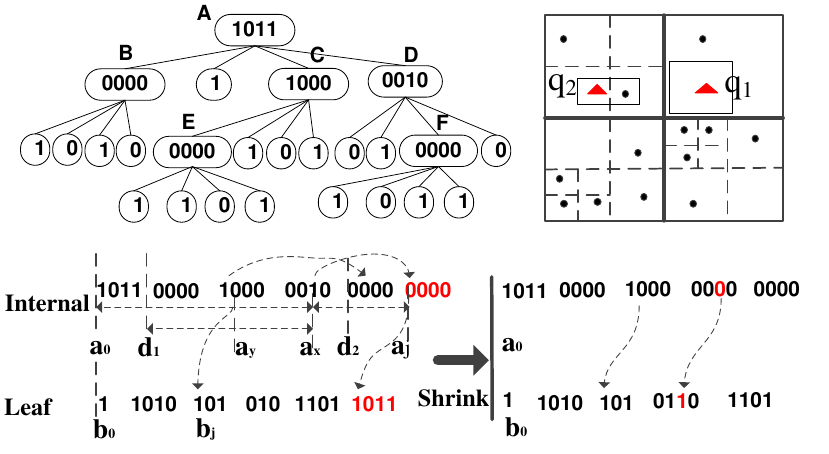}
\caption{sFilter structure (up left), the related data (up right) and the two bit sequences of the sFilter (down).}
\label{fig:sfilter_example}
\vspace{-1em}
\end{figure}

%basic data structure inner node and leaf node
\begin{sloppypar}
Figure~\ref{fig:sfilter_example} gives an example of an sFilter. %The sFilter is a new in-memory index motivated by in-memory indexes~\cite{Radix, rangeFilter}. 
Conceptually, an sFilter is a new in-memory variant of a quadtree that has internal and leaf nodes~\cite{SametBook}. 
Internal nodes are for index navigation, and 
leaf nodes, each has a marker to indicate whether or not there are data items in the node's corresponding region. 
We encode the sFilter into two binary codes and execute queries over this encoding.
\end{sloppypar}

\subsubsection{Binary Encoding of the sFilter}

\begin{sloppypar}
The sFilter is encoded into two long sequences of bits. The first bit-sequence corresponds to internal nodes while the second bit-sequence corresponds to leaf nodes. Notice that in these two binary sequences, no pointers are needed. Each internal node of the sFilter takes four bits, where each bit represents one of the internal node's children. 
These children are encoded in clock-wise order starting from the upper-left corner. 
Each bit value of an internal node determines the type of its corresponding child, i.e., whether the child is internal (a 1 bit) or leaf (a 0 bit). In Figure~\ref{fig:sfilter_example}, the root (internal) node $A$  has binary code $1011$, i.e., 
%(in clock time order) 
it has three of its children being internal nodes, and its second node is leaf. 
The four-bit encodings of all the internal nodes are concatenated  to form the internal-node bit-sequence of the sFilter. The ordering of the internal nodes  in this sequence is based on a breadth-first search (BFS) traversal of the quadtree.  
In contrast, a leaf node only takes one bit, and its bit value indicates whether or not data points exist inside the  spatial quadrant corresponding to the leaf. In Figure~\ref{fig:sfilter_example}, internal node $B$  has four children, and the bit values for $B$'s leaf nodes are $1010$, i.e., the first and third leaf nodes  of $B$ contain data items.
%MO: Check. The previous sentence was horrible!
%Mingjie: thanks, correct
During the BFS on the underlying quad-tree of the sFilter, simultaneously construct the bit-sequences for all the leaf and internal nodes.
%To encode the bit-sequence for all the leaf nodes in an sFilter, 
%during the same BFS on the underlying quad-tree of the sFilter to produce the bit-sequence for the internal nodes, 
%we simultaneously construct the bit-sequence for all the leaf nodes. 
%Thus, we produce two encodings, one for a BFS of the internal nodes and one for a BFS of the leaf nodes.
The sFilter  is encoded into the two binary sequences in Figure~\ref{fig:sfilter_example}. 
The space usage of an sFilter is $O(((4^{d-1}-1)/3) \times 4 +  4^{d-1} )$ Bits, where $O((4^{d-1}-1)/3)$ and $O(4^{d-1})$ are the numbers of internal nodes and leaf nodes, respectively, and $d = o(\log(L))$ is the depth of quadtree and $L$ is the length of the space. 
\end{sloppypar}

%AR i noticed one thing, your sFilter has very high similarity with the pyramid structure proposed by prof Aref, the encoding of nodes is described as colors in the pyramid structure, the pyramid structure is highly used, e.g., used by Amr magdy in tagreed his his work. You may need to differentiate your index from the pyramid structure. 
%Mingjie: the sfilter actually is a quadtree, but it is encoded into binary code and search over the binary code, this where is speedup comes from.  
%AR so the the issue here is that if you check the modifications of incomplete pyramid in Tagreed, it is in fact a quad tree, with similar encoding, this stucutre has been used in multiple recent papers, you may want to mention some difference between them. the main difference i see here is the encoding scheme. the incomplete pyramid uses a different encoding scheme than yours

\subsubsection{Query Processing Using the sFilter}

Consider the internal node $D$ in Figure~\ref{fig:sfilter_example}. $D$'s binary code is $0010$, and the third bit has a value of 1 at memory address $a_x$ of the internal nodes bit sequence.
Thus, this bit refers to $D$'s child $F$ that is also an internal node at address $a_j$. Because  the sFilter has no pointers, we need to compute  $F$'s address $a_j$ from
$a_x$. Observe that the number of bits with value 1 from the start address $a_0$ of the binary code
to $a_x$ 
can be used to compute the address. 
\begin{sloppypar}
\begin{definition}
Let $a$ be the bit sequence that starts at address $a_0$. $\chi(a_0, a_x)$ and $\tau(a_0, a_x)$ are the number of bits with  value 1 and 0, respectively, from addresses
$a_0$ to $a_x$ inclusive. 
\end{definition}
\end{sloppypar}
%In the sFilter, each bit with value 1 represents an internal node. Thus,
%MO: make proper reference to the example.
$\chi(a_0, a_x)$ is  the number of internal nodes up to $a_x$. 
Thus, the address $a_j$ of $F$ is $(a_0+5 \times 4)$ because there are 5 bits with value 1 from $a_0$ to $a_x$. Similarly, if one %children 
child
node is a leaf node, its address is inferred from $\tau(a_0, a_x)$ as follows:
\begin{proposition}
\label{lemma:addressofChildren}
Let $a$ and $b$ be the sFilter's bit sequences for internal and leaf nodes in memory addresses $a_0$ and $b_0$, respectively. 
To access a node's child in memory, we need to compute its address. The address, say $a_j$, of the $x$th child of an internal node at address $a_x$ is computed as follows. 
If the bit value of $a_x$ is 1, then $a_j=a_0+ 4 \times \chi(a_0, a_x)$. If the bit value of $a_x$ is 0, $a_j=b_0+\tau(a_0, a_x)$. 
\end{proposition}

We adopt the following two optimizations to speedup the computation of $\chi(a_0, a_x)$ and $\tau(a_0, a_x)$: (1)~Precomputation and (2)~Set counting. 
Let $d_i$ be the memory address of the first internal node at height (or depth) $i$ of the  underlying quadtree when traversed in BFS order. For example, in Figure~\ref{fig:sfilter_example}, nodes $B$ and $E$ are the first internal nodes in BFS order at depths 1 and 2 of the quadtree, respectively.
For all $i \le$ depth of the underlying quadtree, we precompute $\chi(a_0, d_i)$, e.g., $\chi(a_0, d_1)$ and $\chi(a_0, d_2)$ in Figure~\ref{fig:sfilter_example}. Notice that $d_0 = a_0$ and $\chi(a_0, d_0) = 0$.
Then, address $a_j$ that corresponds to the memory address of the $x$th child of an internal node at address $a_x$ can be computed as follows.
$a_j =a_0+ (\chi(a_0, d_1)+ \chi(d_1,a_x))\times 4$.
$\chi(a_0, d_1)$ is precomputed. Thus, we only need to compute on the fly $\chi(d_1, a_x)$.
Furthermore, evaluating $\chi$ can be optimized by a bit set counting approach, i.e, a lookup table or a sideways addition~\footnote{\url{https://graphics.stanford.edu/~seander/bithacks.html}} that can achieve constant time complexity.

After getting one node's children via Proposition~\ref{lemma:addressofChildren}, we apply Depth-First Search (DFS) over the binary codes of the internal nodes
%(say $a$) 
to answer a spatial range query.
The procedure starts from the first four bits of bit sequence $a$, since these four bits are the root node of the sFilter.  
Then, we check the four
%rectangles
quadrants, say $r_s$,
of the children of the root node, 
and iterate over $r_s$ to find 
the quadrants, say $r_s'$, overlapping the input query range $q_i$.
Next, we continue searching the children of $r_s'$ based on the addresses computed from Proposition~\ref{lemma:addressofChildren}. 
This recursive procedure stops if a leaf node is found with value 1, or if all internal nodes are visited. 
For example, consider range query $q_2$ in Figure~\ref{fig:sfilter_example}. We start at the root node $A$ (with bit value $1011$). Query $q_2$ is located inside the northwestern (NW) quadrant of $A$. Because the related bit value for this quadrant is 1, it indicates an internal node type and it refers to child node $B$. Node $B$'s memory address is computed by $a_0+1 \times 4$ because only one non-leaf node ($A$) is before $B$. $B$'s related bit value is $0000$, i.e., $B$ contains four leaf nodes. 
The procedure continues until finding one leaf node of $B$, mainly the southeastern child leaf node, with value 1 that overlaps the query, and thus returns true.

\begin{algorithm}[t]\small
\KwIn{
$LocationRDD$: Distributed/indexed spatial data, \\
~~~~~~~~~~~$Q$: Input set of spatial range queries
}
\KwOut{$R$: Results of the spatial queries}
Var index $\leftarrow$  $LocationRDD.index$ //get global index with embedded sFilters\\ 
Var qRDD $\leftarrow$ partition($Q$,index) // Distribute in parallel the input spatial queries using the global index\\
Var update\_sFilter $\leftarrow$ //function for updating the sFilter in each worker\\
\{
\\
   \For{each query $q_i$ in this worker}
   {
        \If {query $q_i$'s return result is empty}
        {
           insert($q_i$, sFilter) // adapt sFilter given $q_i$
        }
   }

   \If{sFilter.space $> \alpha$}
   {
       shrink(sFilter) // shrink the sFilter to save space
   }

\}

$R$ $\leftarrow$ $LocationRDD$.sjoin(qRDD)(update\_sFilter) //execute spatial join and update sFilter in workers \\
Var sFilters $\leftarrow$ $LocationRDD$.collect\_sFilter() //collect sFilter from workers\\
mergesFilters(sFilters, index) // update sfilter in global index\\
return $R$
\caption{\small{Update sFilter in LocationSpark}}
\label{alg:sfilterForSpatialJoin}
\end{algorithm}

\subsection{sFilter in LocationSpark}

Since the depth of the sFilter affects query performance, it is impractical to use only one sFilter in a distributed setting. 
We thus embed multiple sFilters into the global and local spatial indexes in~\system. In the master node, separate sFilters are placed into the different branches of the global index, where the role of each sFilter is to locally answer the query for the specific branch it is in.
In the local computation nodes, an sFilter is built and it adapts its structure based on data updates and changes in query patterns. 
%Note that internal nodes of sFilter in local node is implemented with pointers to their children as standard Quadtree, since it is efficient to split and merge nodes of sFilter.
%WGA: I do not see the significance of the above sentence.
%Mingie: Ok

\subsubsection{ Spatial Query Processing Using the sFilter}

Algorithm~\ref{alg:sfilterForSpatialJoin} gives the procedure for performing the spatial range join using the sFilter. Initially, the outer (queries) table  is partitioned according to the global index. The global index identifies the overlapping data partitions for each query $q$. Then, the sFilter tells which partitions contain data that overlap the query range (Line~2 of the algorithm).
After performing the spatial range join (Line~14), the master node fetches the updated sFilter from each data worker, 
and refreshes the existing sFilters in the master node
(Lines~15-16). Lines~2-13 update the sFilter of each worker (as in Figure~\ref{fig:system_overview}). 

The sFilter can
improve the $k$NN search and $k$NN join because  they also depend on  spatial range search. 
Moreover, their query results may enhance the sFilter by lowering the false positive errors as illustrated below.

\subsubsection{Query-aware Adaptivity of the sFilter}
\label{section:sfilterAdaptive}
The build and update operations of the sFilter are first executed at the local workers in parallel. Then, the updated sFilters are propagated to the master node. 

%The procedure to build sFilter contains two stages. 
%The first stage is to build 
The initial sFilter is  built from a temporary local  quadtree~\cite{SametBook} in each partition.
Then, the sFilter is adapted based on the query results. For example, consider Query $q_1$ in Figure~\ref{fig:sfilter_example}. Initially, the sFilter reports that there is data for $q_1$ in the partitions. When $q_1$ visits the  related data partitions, it finds that there are actually no data points overlapping with $q_1$ in the partitions, i.e., a false-positive (+ve) error. 
Thus, we mark the quadrants precisely covered by $q_1$ in the sFilter as empty, and hence reduce the false positive errors if queries visit the marked quadrants again. Function \texttt{insert} in Algorithm~\ref{alg:sfilterForSpatialJoin} recursively splits the quadrants covered by the empty query, and marks these generated quadrants as empty.
%until the quadrant of sFilter coaligns with the boundary of input query. 
After each local sFilter is updated in each worker, these updates are reflected into the master node. The compact encoding of the sFilter saves the communication cost between the workers and the master. 

However, the query performance of the sFilter degrades as the size of the index increases. Function \texttt{shrink} in Algorithm~\ref{alg:sfilterForSpatialJoin} merges some branches of the sFilter at the price of increasing false +ve errors. For example, one can shrink internal node $F$ in Figure~\ref{fig:sfilter_example} into a leaf node, and updating its bit value to 1, although one quadrant of $F$ does not contain data. Therefore, we might track the visit frequencies of the internal nodes, and merge internal nodes with low visiting frequency. Then, some well-known data caching policies, e.g., LRU or MRU, can be used. However, the overhead to track the visit frequencies is expensive. In our implementation, we adopt a simple bottom-up approach. We start merging the nodes from the lower levels of the index to the higher levels until the space constraint is met. In Figure~\ref{fig:sfilter_example}, we shrink the sFilter from internal node $F$, and replace it by a leaf node, and update its binary code to $1$. $F$'s leaf children are removed. The experimental results show that this approach increases the false +ve errors, but enhances the overall query performance.

\section{Performance Study}
\label{section:evaluation}

\system is implemented on top of Resilient Distributed Datasets (RDDs); these key components of Spark are 
 fault-tolerant collections of elements that can be operated on in parallel. 
%RDDs, which are the distributed memory abstraction in Spark. 
\system is a library of Spark and provides the Class LocationRDD for spatial 
%MO:Why the reference?
%Mingjie: because the API is attached in that tech report
operations~\cite{LocationSparkTechReport}.  Statistics are maintained at the
driver program of Spark, and the execution plans are generated at the driver. Local spatial indexes are persisted
in the RDD data partitions, while the global index is realized by extending the interface of the RDD data partitioner. The data tuples and related spatial indexes are encapsulated into the RDD data partitions. Thus, Spark's fault tolerance naturally applies in~\system. The spatial indexes are immutable and are implemented based on the path copy approaches. Thus, each updated version of the spatial index can be persisted into disk for fault tolerance. This enables the recovery of a local index from disk in case of failure in a worker. The Spark cluster is managed by YARN, and a failure in the master nodes is detected and managed by ZooKeeper. In case of master node failure, the lost master node is evicted and a standby node is chosen to recover the master. As a result, the global index and the sFilter in the master node are recoverable. Finally, the built spatial index data can be stored into disk, and enable further data analysis without additional data repartitioning or indexing.  \system is  open-source, and can be downloaded from \url{https://github.com/merlintang/SpatialSpark}.
%AR i think the previous statement is not very important 
%AR in here you mention several terms not cited and not mentioned before 
%AR i think the main point in the previous paragraph is that you want to describe how location spark is implemented on top of spark. and you may focus on the following points: (1) location spark is an API on top of spark that can be seamlessly integrated with any spark installation (2) you provide an extension of the scala query language to support your spatial queries. (3) all your extensions inherit fault tolerance from spark.  
%AR i believe the location of the local and global indexes should not be stated in the experimental section 
%AR maybe you can move the fault tolerance discussion into a separate section

\begin{figure*}
        \centering
        \begin{subfigure}[b]{0.21\textwidth}
                \includegraphics[width=\columnwidth]{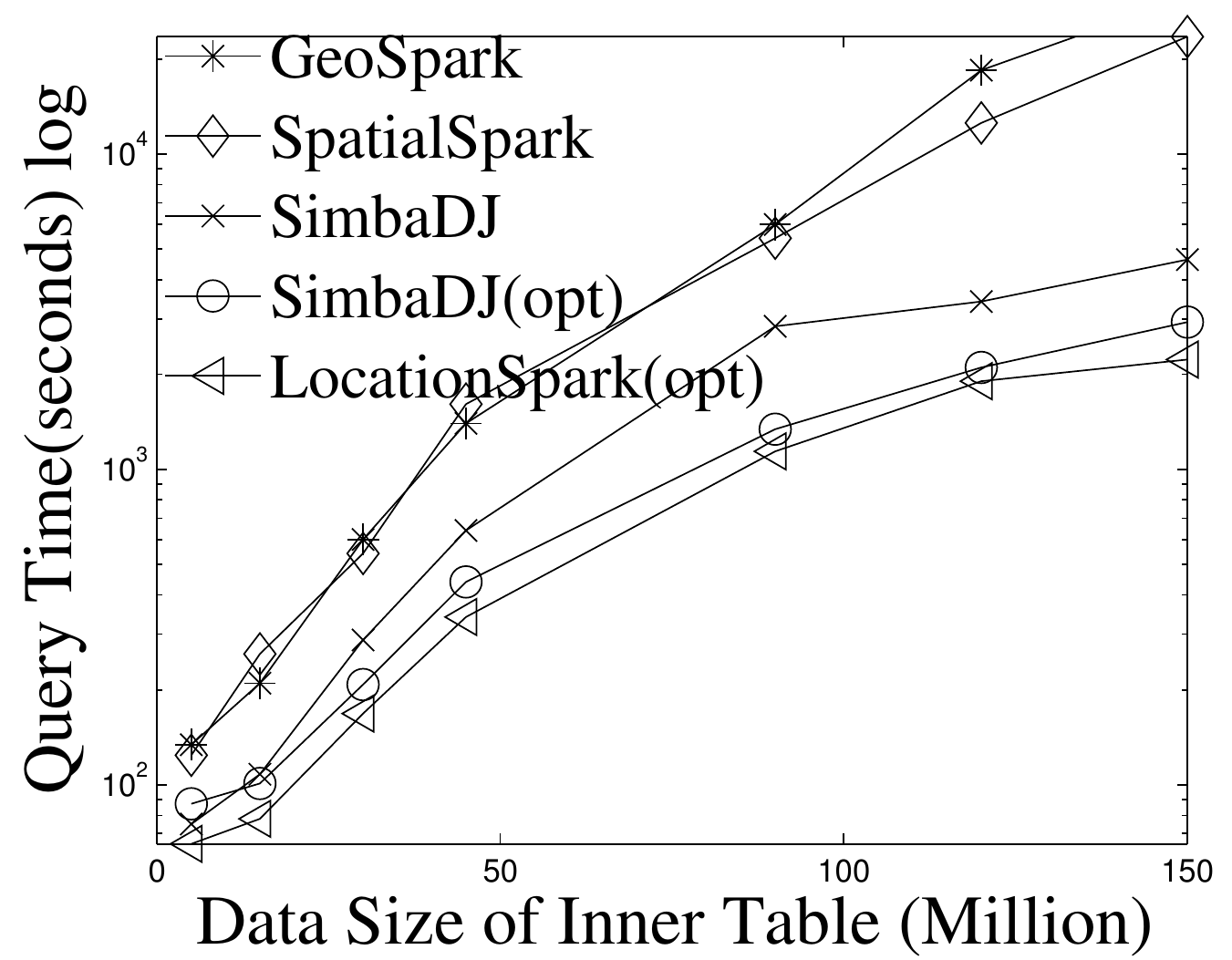}
                \caption{Twitter}
                \label{fig:sjoin_effect_twitter}
        \end{subfigure}%
        \begin{subfigure}[b]{0.21\textwidth}
               \includegraphics[width=\columnwidth]{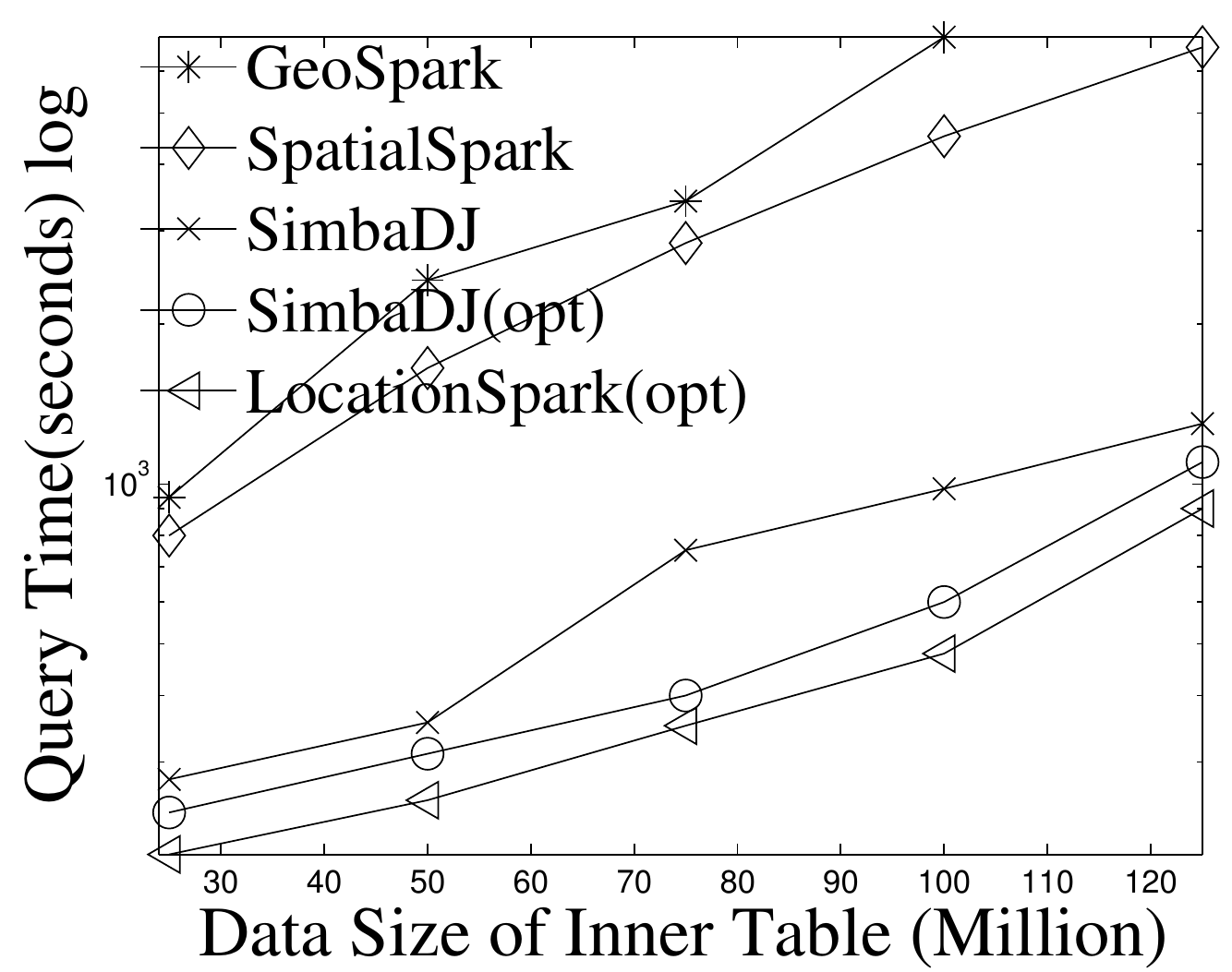}
                \caption{OSMP}
                \label{fig:sjoin_effect_osmp}
        \end{subfigure}
        \begin{subfigure}[b]{0.21\textwidth}
               \includegraphics[width=\columnwidth]{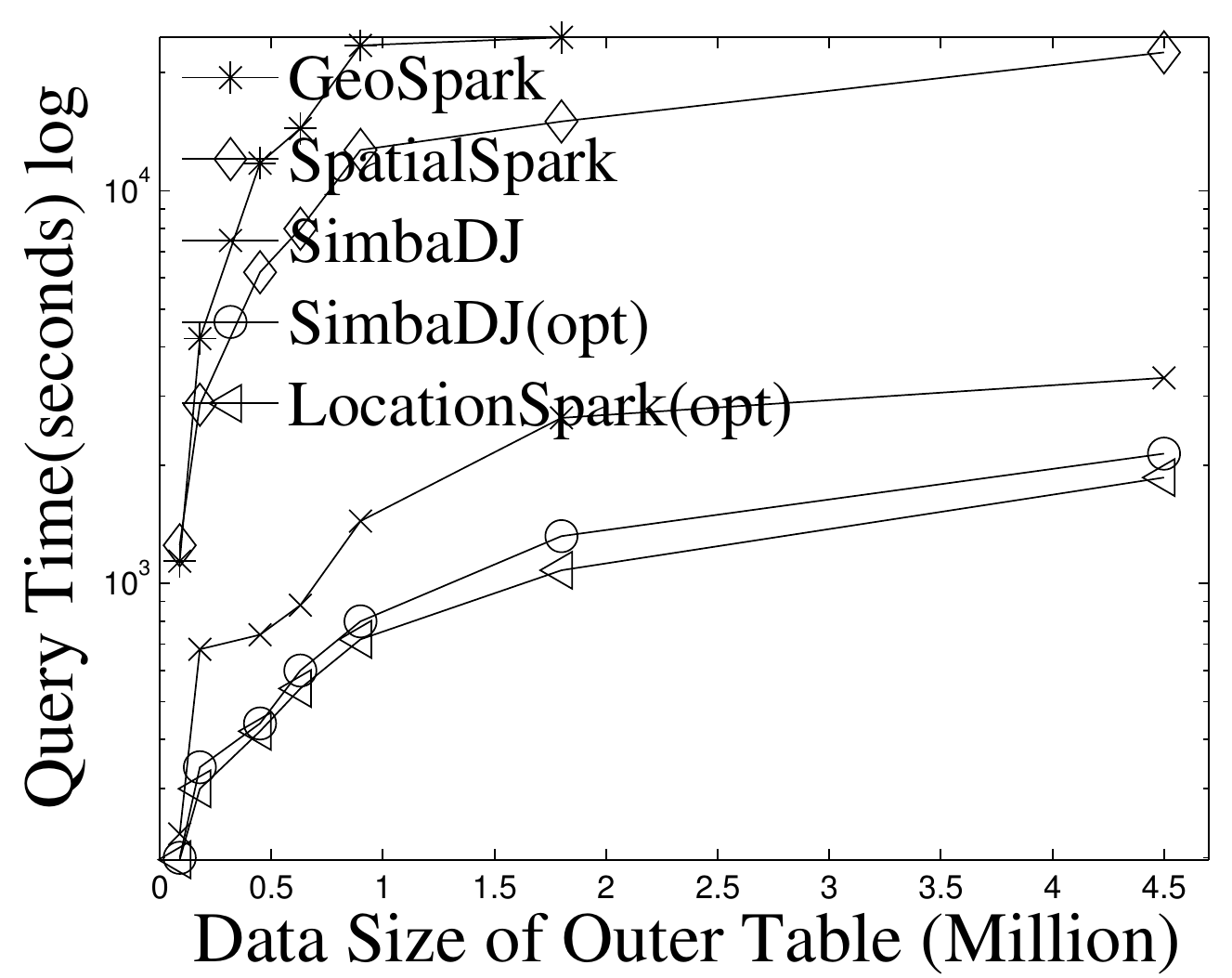}
                \caption{Twitter}
                \label{fig:sjoin_effect_numberofboxes_twitter}
        \end{subfigure}
        \begin{subfigure}[b]{0.21\textwidth}
               \includegraphics[width=\columnwidth]{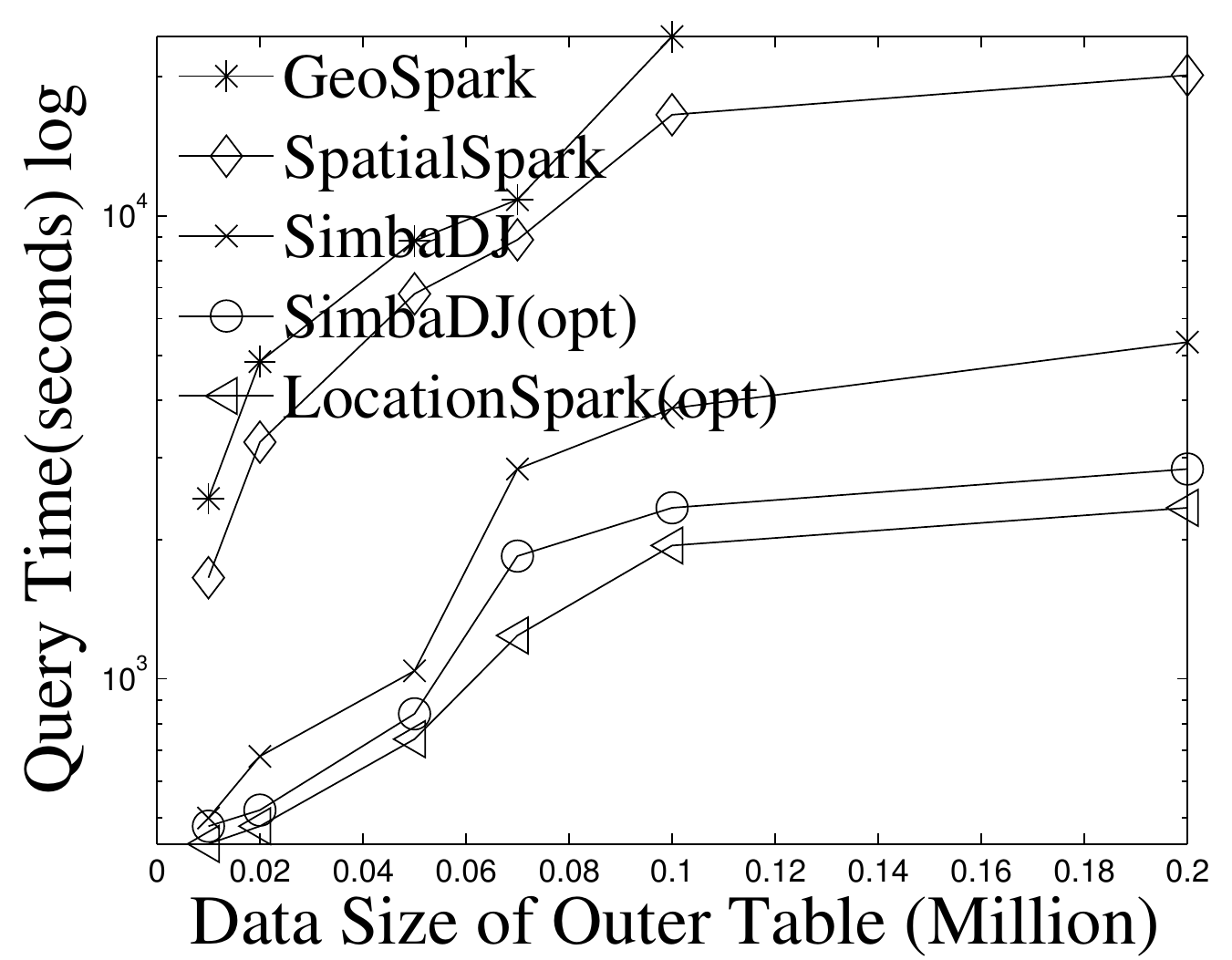}
                \caption{OSMP}
                \label{fig:sjoin_effect_numberofboxes_osmp}
        \end{subfigure}
        \caption{The performance of spatial range join }
        \label{fig:sjoin_test}
        \vspace{-1em}
\end{figure*}

\subsection{Experimental Setup}

%\textbf{Datasets.} 
Experiments are conducted on two datasets. \textbf{Twitter}: 1.5 Billion Tweets (around 250GB) are collected over a period of nearly 20 months (from January 2013 to July 2014) and is restricted to the USA spatial region. The format of a tweet is:  identifier, timestamp, longitude-latitude coordinates, and text. \textbf{OSMP}: is shared by the authors of SpatialHadoop~\cite{spatialhadoop}.
OSMP represents the map features of the whole world, where each spatial object is identified by its coordinates (longitude,  latitude) and an object ID. It contains 1.7 Billion points with a total size of 62.3GB. We generate two types of queries. (1)~Uniformly distributed (USA, for short): We uniformly sample data points from the corresponding dataset and generate spatial queries from the samples. These are the default queries in our experiments. (2)~Skewed spatial queries: These are synthesized around specific spatial areas, e.g., Chicago, San Francisco, New York (CHI, SF, NY, correspondingly, for short). The spatial queries and data points are the outer table $Q$ and the inner table $D$ for the experimental studies of the spatial range and $k$NN joins   presented below.   

%AR i think the previous sentence needs some rewrite especially the the synthetic dataset part 

%\textbf{Comparisons.} 
%We carried out extensive experiments to evaluate the performance of~\system. In particular, the performance of
Our study compares \system with the following:
(1)~\textbf{GeoSpark}~\cite{GeoSpark} uses  ideas from SpatialHadoop but is implemented over Spark. (2)~\textbf{SpatialSpark}~\cite{SpatialSpark} performs partition-based spatial joins. (3)~\textbf{Magellan}~\cite{Magellan} is developed based on Spark's dataframes  to take advantage from Spark SQL's plan optimizer. However, Magellan does not have spatial indexing.
(4)~\textbf{State-of-art $k$NN-join}: Since none of the three systems support $k$NN join, we compare~\system with a state-of-art $k$NN-join approach (PGBJ~\cite{PGBJ}) that is provided by PGBJ's authors. 
(5)~\textbf{Simba}~\cite{Simba} is a spatial computation system based on Spark SQL with spatial distance join and $k$NN-join operator. We also modified Simba with the developed techniques (e.g., query scheduler and sFilter) inside, the optimized Simba is called \textbf{Simba(opt)}. 
%MO: You mean LocationSpark(opt) with sFilter and  \textbf{LocationSpark} without?
%Mingjie: yes it is. 
(6) \textbf{LocationSpark(opt)} and \textbf{LocationSpark} refers to the query scheduler and sFilter is applied or not, respectively. 

We use a cluster of six physical nodes  Hathi~\footnote{https://www.rcac.purdue.edu/compute/hathi/}. 
that consists of Dell compute nodes with two 8-core Intel E5-2650v2 CPUs, 32 GB of memory, and 48TB of local storage per node. Meanwhile, in order to test the scalability of system, we setup one Hadoop cluster (with Hortonworks data paltform 2.5) on the the Amazon EC2 with 48 nodes, each node has an Intel Xeon E5-2666 v3 (Haswell) and 8GB of memory. Spark 1.6.3 is used with YARN cluster resource management.  Performance is measured by the average query execution time.
\begin{table}[htbp]
\centering
\scalebox{0.8}{
\begin{tabular}{llll} \hline
Dataset
& System
&\vtop{\hbox{\strut Query}\hbox{\strut time(ms)}}
&\vtop{\hbox{\strut Index}\hbox{\strut build time(s)}}\\
\hline
\multirow{5}{*}{\centering Twitter}
&LocationSpark(R-tree)&390&32\\
&LocationSpark(Qtree)&\textbf{301}&16\\
&Magellan&15093&/\\
&SpatialSpark&16874&35\\
&SpatialSpark(no-index)&14741&/\\
&GeoSpark&4321&45\\
&Simba&1231&34\\
&Simba (opt)&430&35\\
\hline
\multirow{5}{*}{\centering OSMP}
&LocationSpark(R-tree)&1212&67\\
&LocationSpark(Qtree)&\textbf{734}&18\\
&Magellan&41291 &/\\
&SpatialSpark&24189&64\\
&SpatialSpark(no-index)&17210&/\\
&GeoSpark&4781&87\\
&Simba&1345&68\\
&Simba(opt)&876&68\\
\hline
\end{tabular}
}
\caption{Performance of the spatial range search}
\label{tab:spatial_search_comparision}
\end{table}

\begin{figure}
        \centering
        \begin{subfigure}[b]{0.19\textwidth}
                \includegraphics[width=\columnwidth]{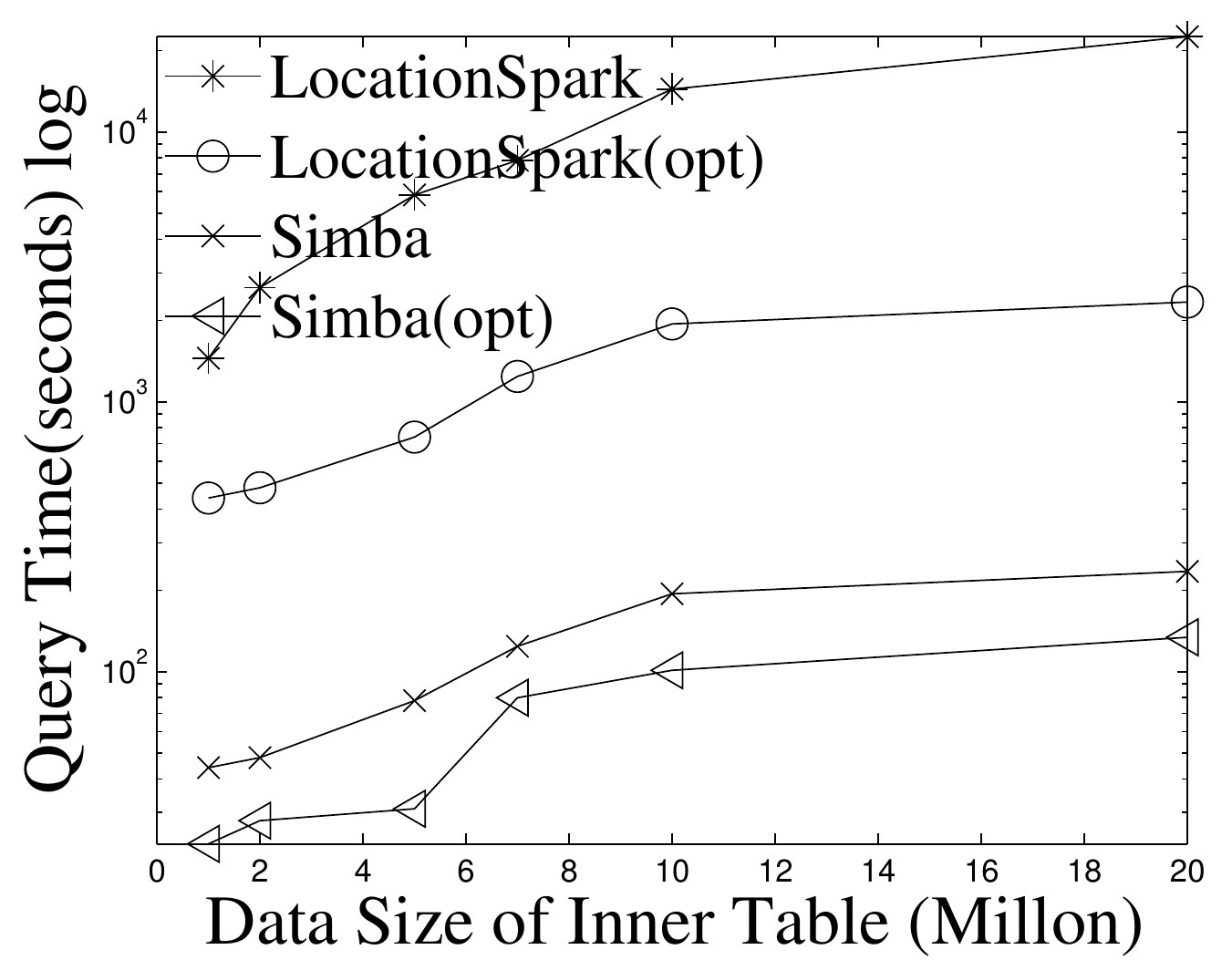}
                \caption{Twitter}
                \label{fig:knnjoin_datasizetwitter}
        \end{subfigure}%
        \begin{subfigure}[b]{0.19\textwidth}
               \includegraphics[width=\columnwidth]{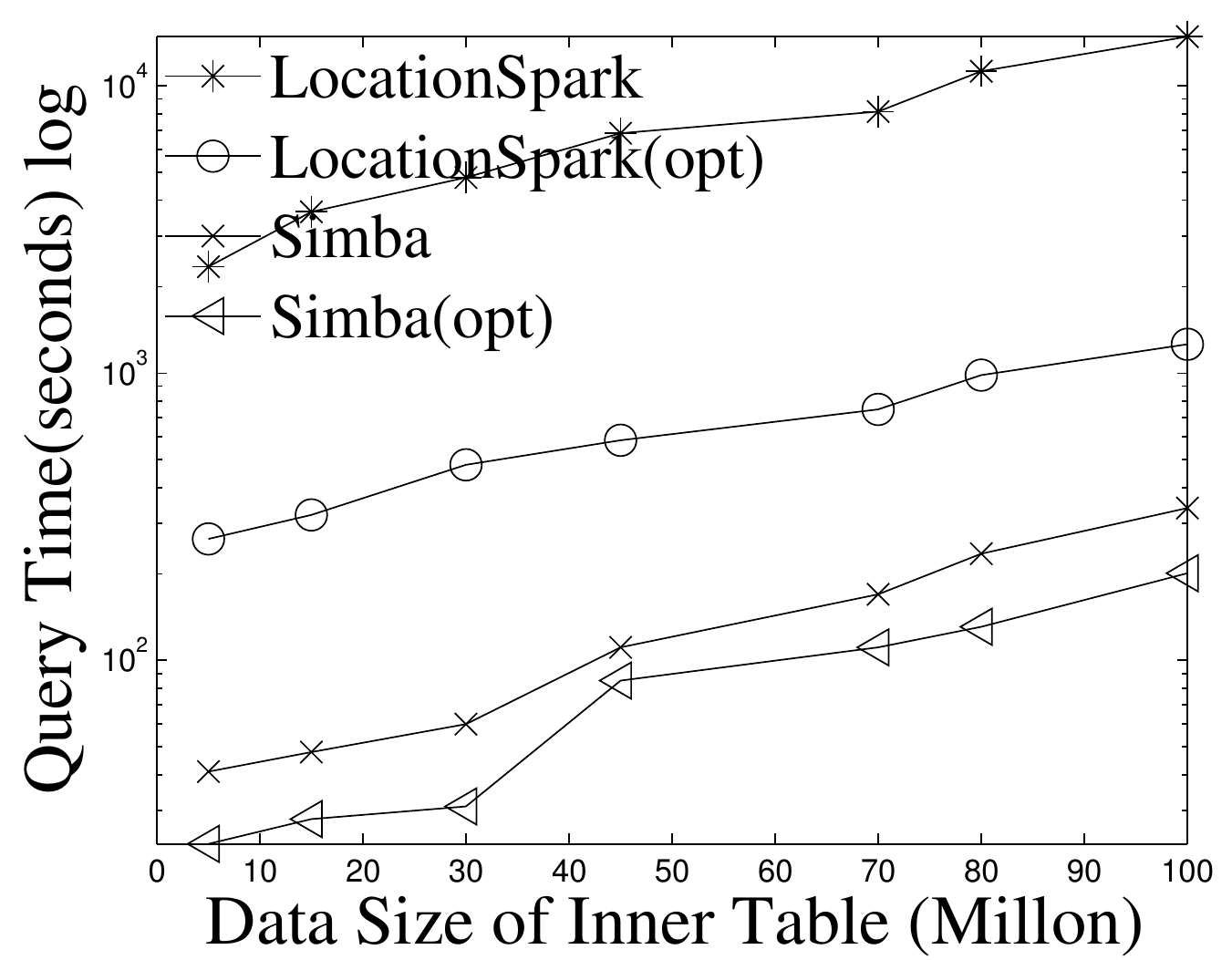}
                \caption{OSMP}
                \label{fig:knnjoin_datasize_osmp}
        \end{subfigure}
        \caption{Performance of $k$NN join by increasing the number of data points}
        \label{fig:knnjoin_datasize}
        \vspace{-1em}
\end{figure}

\subsection{Spatial Range Search and Join}

Table~\ref{tab:spatial_search_comparision} summarizes the spatial range search and spatial index build time by the various approaches. For a fair comparison, we cache the indexed data into memory, and record the spatial range query processing time. From Table~\ref{tab:spatial_search_comparision}, we observe the following: 
(1)~\system is 50 times better than  
Magellan on query execution time for the two tables, mainly because the spatial index (e.g., Global and Local index) of~\system can avoid visiting unnecessary data partitions.
(2)~\system with different local indexes, e.g., the R-tree and Quadtree, outperforms SpatialSpark. The speedup is around 50 times, since SpatialSpark (without index) has to scan all the data partitions.
SpatialSpark (with index) stores the global indexes into disk, and finds data partitions by scanning the global index in disk. This incurs extra I/O overhead. Also, the local index is not utilized during local searching. 
(3)~\system is around 10 times faster than GeoSpark in spatial range search execution time because GeoSpark does not utilize the built global indexes and scans all  data partitions. 
(4)~The local index with Quadtree for~\system achieves superior performance over the R-tree one in term of index construction and query execution time as discussed in Section~\ref{section:locatlPlan}. 
(5) The index build time among the three systems is comparable because they all scan the data points, which is the dominant factor, and then build the index in memory. 
(6)~\system achieve 5 times speedup against with Simba, since sFilter can reduce redundant data partitions searching. This is also observed from the Simba(opt) (i.e., with sFilter) can achieve comparable performance with ~\system.      

Performance results (mainly, the execution times of the spatial range join) are listed in Figure~\ref{fig:sjoin_test}.  For fair comparison, the runtime 
%is counted as end to end, which 
includes the time to initiate the job, build indexes, execute the join query, and save results into HDFS. Note that Simba does not support spatial range join, it only provides the spatial distance join, where each query is rectangle or circle with the same size.  
%as the number of data sizes in $D$ and $Q$ increase. 
Performance results for Magellan are not shown because it performs a Cartesian product and hence has the worst execution time. 
Figures~\ref{fig:sjoin_effect_twitter} and~\ref{fig:sjoin_effect_osmp} present the results by varying the data sizes of $D$ (the inner table) from 25 million to 150 million, while keeping the size of $Q$ (the outer table) to 0.5 million.
The execution time of GeoSpark shows quadratic increase as the data size increases. GeoSpark's running time is almost 3 hrs when the data size is 150 million, which is extremely slow. SpatialSpark shows
similar trends. The reason is that both GeoSpark and SpatialSpark suffer from 
(1)~the spatial skew issue where some workers process more data and take longer time to finish. 
(2)~the local execution plan based on the R-tree and the Grid is slow. 
(3)~processing of queries go to data partitions that do not contribute to the final results. 
\system with the optimized query plans and the sFilter outperforms the two other systems by an order of magnitude. 
Also, we study the effect of the outer table size on performance. Figures~\ref{fig:sjoin_effect_numberofboxes_twitter} and~\ref{fig:sjoin_effect_numberofboxes_osmp} give the run time, and demonstrate that \system  is 10 times faster than the other two systems. From~Figure~\ref{fig:sjoin_test}, we also observe Simba outperform GeoSpark and SpatialSpark more than one order of magnitude, however, Simba also suffer from the query skew issues and degrade the performance quickly, because Simba has to duplicate each data point into multiple data partitions if its spatial overlapping with query rectangles. Naturally, the bigger spatial query rectangle, the more data points to be duplicated. This brings redundant network and computation costs. Thus, we place the query scheduler and sFilter into the physical execution part of Simba (e.g., modifying the RDKspark). From the experimental result, the performance of Simba is improved dramatically. This also proves the proposed approaches in this work could be used for improve most in-memory distributed spatial query processing systems.

\subsection{Performance of kNN Search and Join}

Performance of $k$NN search  is listed in Table~\ref{tab:comparisionwithbaseline_knnsearch}. \system outperforms GeoSpark by an order of magnitude. 
GeoSpark broadcasts the query points to each data partition, and accesses each data partition to get the $k$NN set for the query.
Then, GeoSpark collects the local results from each partition, then sorts the tuples based on the distance to query point of $k$NN. This is prohibitively expensive, and results in large execution time. \system only searches for data partitions that contribute to the $k$NN query point based on the global and local spatial indexes and the sFilter.
It avoids redundant computations and unnecessary network communication for irrelevant data partitions. 

For $k$NN join,
Table~\ref{tab:comparision_knnjoin} presents the performance results when  varying $k$ on the Twitter and OSMP datasets.
In terms of runtime, \system with optimized
query plans and with the sFilter always performs the best. \system without any optimizations gives better performance than that of PGBJ. The reason is due to having in-memory computations and  avoiding expensive disk I/O when compared to MapReduce jobs.
Furthermore, \system with optimization shows around 10 times speedup over PGBJ, because the optimized plan migrates and splits the skewed query regions. 

We test the performance of the $k$NN join operator when increasing the number of data points while having the number of queries fixed to 1 million around the Chicago area. The results are illustrated in Figure~\ref{fig:knnjoin_datasize}. Observe that \system with optimizations performs an order of magnitude better than the basic approach. 
The reason is that the optimized query plan  identifies and repartitions the skewed partitions. In this experiment, the top five slowest tasks in~\system without optimization take around 33 minutes, while more than 75\% tasks only take less than 30 seconds.
On the other hand, with an optimized query plan, the top five slowest tasks take less than 4 minutes. This directly enhances the execution time.  

More interesting, we observe Simba outperforms ~\system around 10 times for the $k$NN join operation. This speedup is achieved by implementing the $k$NN join via a sampling technique in Simba. More specifically, Simba at first samples certain amount of query points and computes the $k$NN join bound, s.t. pruning data partitions which do not contribute $k$NN join results. The computed $k$NN bound could be very tight when the $k$ is small and query points are balanced distributed. More details can be found in the Section 6.3.3 of Simba. However, the computed $k$NN join bound have to do spatial overlapping checking with data points, then duplicate query points to the overlapping data partitions as well. As a result, Simba would suffer from the query skew where certain data partitions are overwhelmed with query points. Furthermore, the bigger $k$ also brings redundant network communication because the $k$NN join bound would be bigger as well. Similar as spatial join, we placed spatial query scheduler and sFilter inside the physical execution part of Simba. The query scheduler can detect the skew data partitions, as well as sFilter to remove data partitions which overlapping with $k$NN join bound but fail to contribute final result. Therefore, the performance of Simba is further improved, e.g., more than three times in this experimental setting.    

\begin{table}[htbp]
\centering
\scalebox{0.80}{
\begin{tabular}{lllll} \hline
Dataset
& System
&\vtop{\hbox{\strut k=10}}
&\vtop{\hbox{\strut k=20}}
&\vtop{\hbox{\strut k=30}}
\\
\hline
\multirow{3}{*}{\centering Twitter}
&LocationSpark(R-tree)&81&82&83\\
&LocationSpark(Q-tree)&74&75&74\\
&GeoSpark&1334&1813&1821\\
&Simba&420&430&440\\
&Simba(opt)&80&83&87\\
\hline
\multirow{3}{*}{\centering OSMP}
&LocationSpark(R-tree)&183&184&184\\
&LocationSpark(Q-tree)&73&73&74\\
&GeoSpark&4781&4947&4984\\
&Simba&330&340&356\\
&Simba(opt)&167&186&190\\
\hline
\end{tabular}
}
\caption{Runtime (in microseconds) of $k$NN search}
\label{tab:comparisionwithbaseline_knnsearch}
\end{table}
%when the dataset size for Twitter, and OSMp is set to 15 million and 50 million tuples, respectively.

\begin{table}[htbp]
\centering
\scalebox{0.80}{
\begin{tabular}{lllll} \hline
Dataset
& System
&\vtop{\hbox{\strut k=50}}
&\vtop{\hbox{\strut k=100}}
&\vtop{\hbox{\strut k=150}}
\\
\hline
\multirow{3}{*}{\centering Twitter}
&LocationSpark(Q-tree)&340&745&1231\\
&LocationSpark(Opt)&165&220&230\\
&PGBJ&3422&3549&3544\\
&Simba&40&44&48\\
&Simba(Opt)&21&22&31\\
\hline
\multirow{3}{*}{\centering OSMP}
&LocationSpark(Q-tree)&547&1241&1544\\
&LocationSpark(Opt)&260&300&340\\
&PGBJ&5588&5612&5668\\
&Simba&51&55&61\\
&Simba(Opt)&23&26&28
\\
\hline
\end{tabular}
}
\caption{Runtime (in seconds) of $k$NN join}
\label{tab:comparision_knnjoin}
\end{table}
%when the dataset size for Twitter, and OSMp is set to 15 million and 50 million tuples, respectively. 

\subsection{Effect of Query Distribution}
\begin{figure}
        \centering
        \begin{subfigure}[b]{0.20\textwidth}
                \includegraphics[width=\columnwidth]{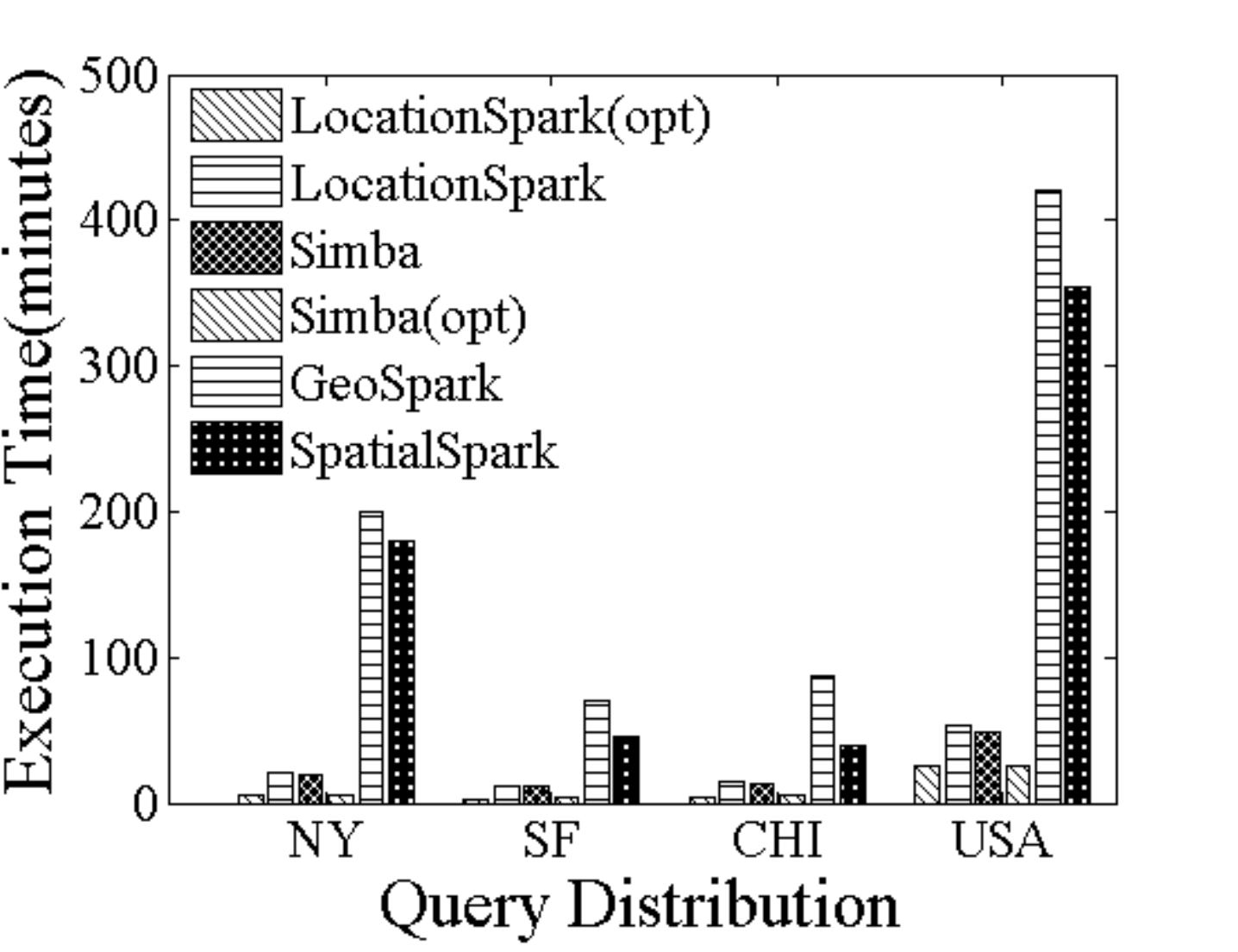}
                \caption{Twitter}
                \label{fig:plan_twitter}
        \end{subfigure}%
        \begin{subfigure}[b]{0.20\textwidth}
               \includegraphics[width=\columnwidth]{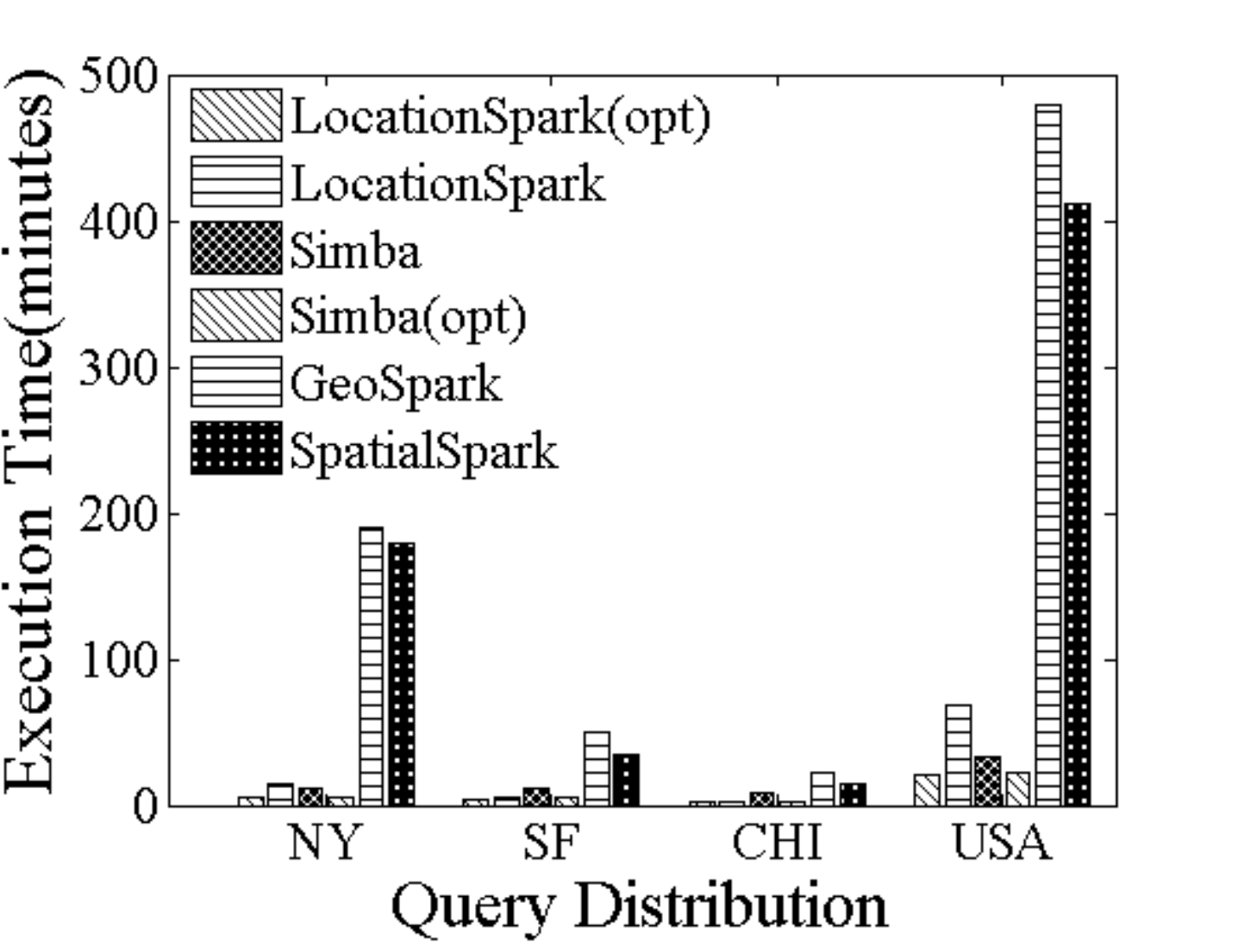}
                \caption{OSMP}
                \label{fig:plan_osmp}
        \end{subfigure}
        \caption{Performance of spatial range join on various query distributions}
        \label{fig:effectOFplan}
        \vspace{-1em}
\end{figure}

We study the performance 
under various query distributions. As illustrated before, the query execution plan is the most effective factor in distributed spatial computing.
From the experimental results for spatial range join and $k$NN join above, we already observe that the system with optimization achieves much better performance over the unoptimized versions.
In this experiment, we study the performance of the optimized query scheduling plan in~\system under various query distributions. The performance of the spatial range join operator over query set  $Q$ (the outer table) and dataset $D$ (the inner table) is used as the benchmark. 
The number of tuples for $D$ is fixed as 15 million and 50 million for Twitter and OSMP, respectively,
while the size of $Q$ is 0.5 million, and each query in $Q$ is generated from different spatial regions, e.g., CHI, SF, NY and USA. We do not plot the runtime of Magellan on spatial join, as it uses Cartesian join, and hence has the worst performance.  Figure~\ref{fig:effectOFplan} gives the  execution runtimes for the spatial range join operators in different spatial regions.
%, where~\system(opt) means the system with the optimized query plan. 
From Figure~\ref{fig:effectOFplan}, GeoSpark performs the worst, followed by SpatialSpark and then~\system. \system and Simba(opt) with the optimized query plan achieves an order of magnitude speedup over
GeoSpark and SpatialSpark in terms of execution time. \system with optimized plans achieves  more than 10 times speedup over \system without the optimized plans for the skewed spatial queries. Meanwhile, Simba could be further improved more than five times comparing with system without optimization.

\subsection{Effect of the sFilter}%AR Effect of the sFilter

%MO: The setting for this experiment are not clear. What are the queries? 
%MO: How come the query execution time is much smaller than the previous numbers shown in Figure 7?
%Mingjie: this is the time on the single node
%MO: Also, do the previous figures use sFilter? 
%Mingjie: yes it already used

\begin{figure}
        \centering
        \begin{subfigure}[b]{0.19\textwidth}
                \includegraphics[width=\columnwidth]{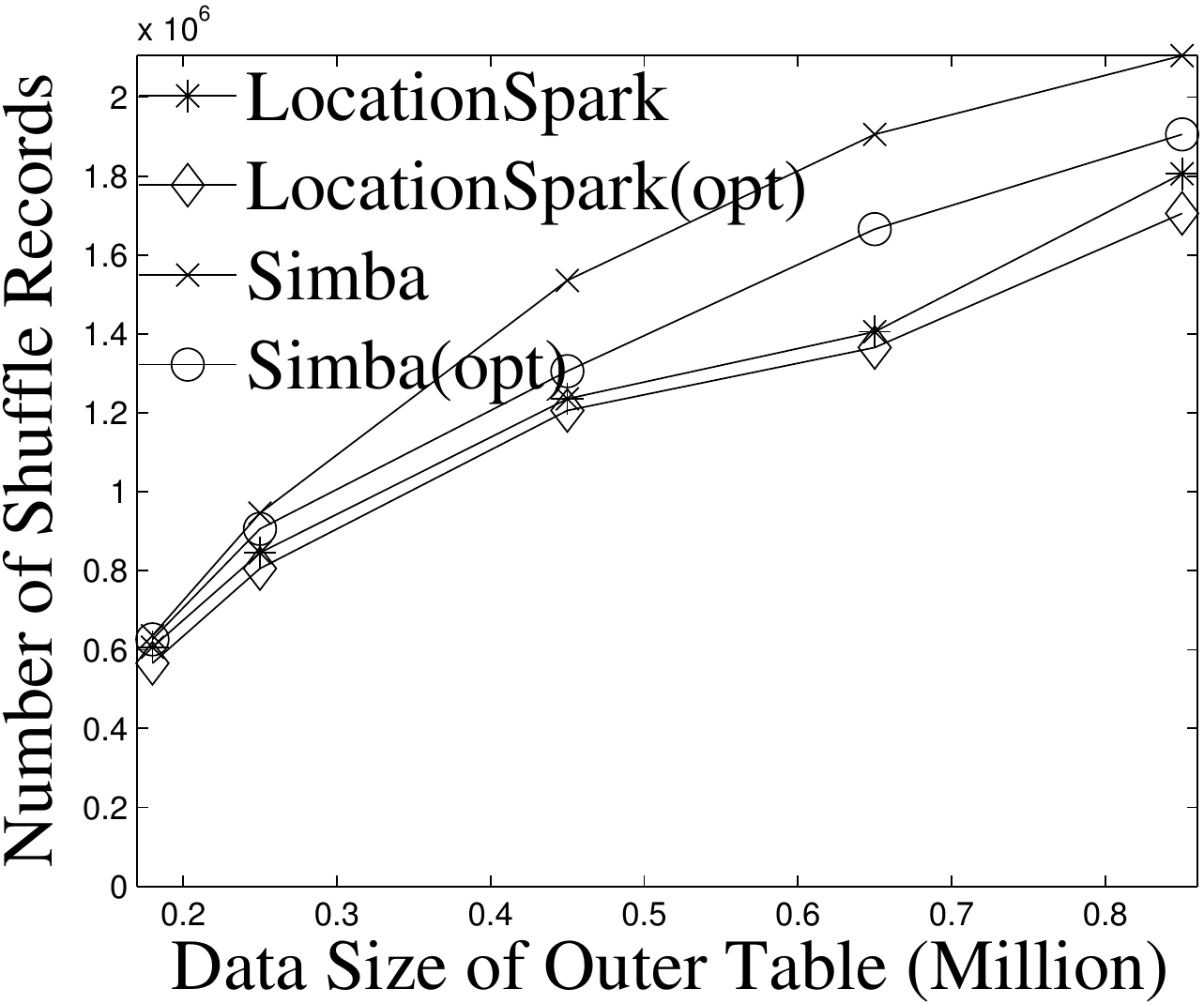}
                \caption{Spatial range join}
                \label{fig:spatialjoin_twitter_sfilter}
        \end{subfigure}%
        \begin{subfigure}[b]{0.19\textwidth}
               \includegraphics[width=\columnwidth]{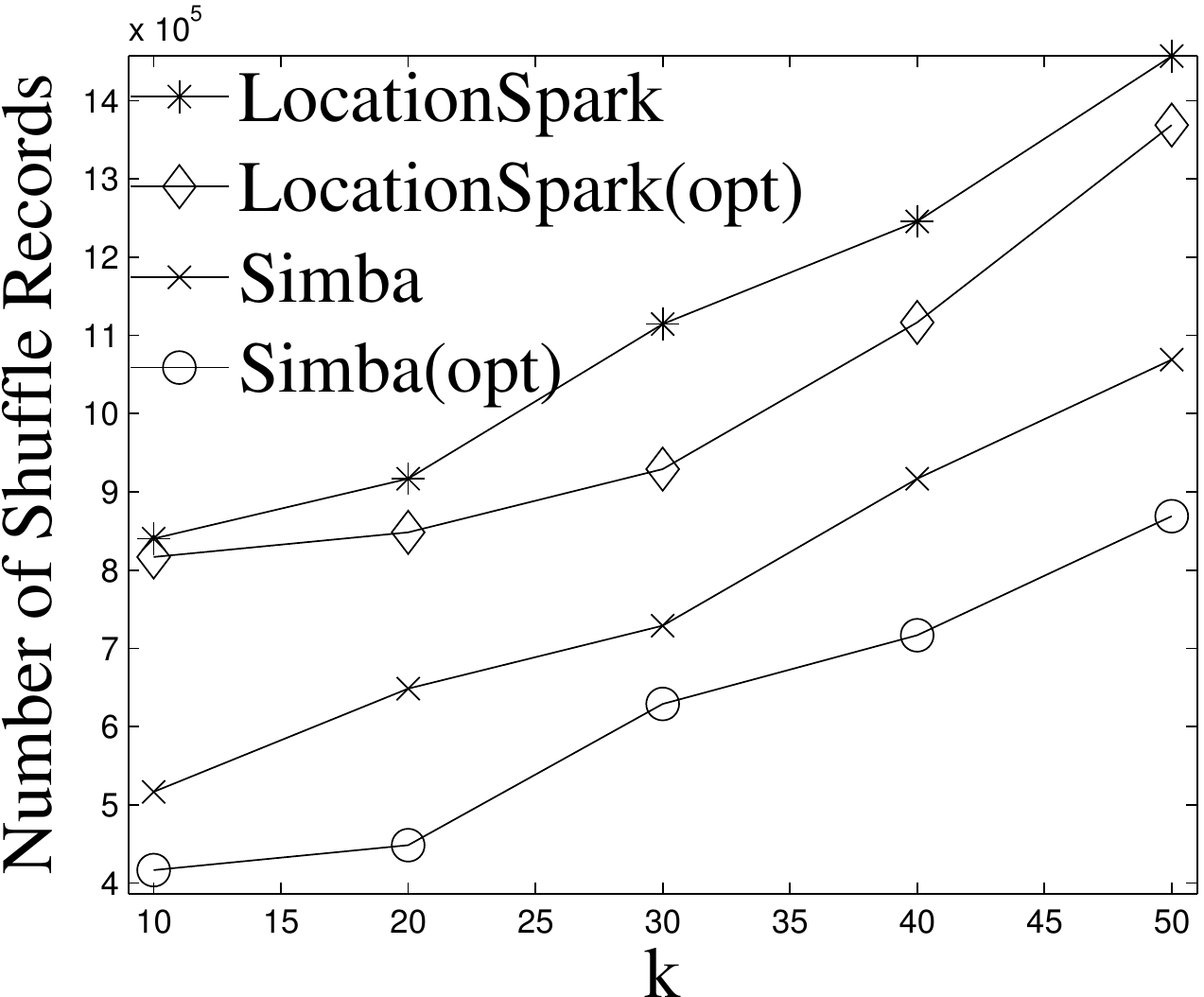}
                \caption{$k$NN join} \label{fig:knnjoin_twitter_sfilter}
        \end{subfigure}
        \caption{The effect of the sFilter on reducing the shuffle cost}
        \label{fig:effect_of_sfilter}
        \vspace{-1em}
\end{figure}

In this experiment, we measure the spatial range query processing time, the index construction
time, the false positive ratio and the space usage of the sFilter on a single machine.
Table~\ref{tab:performanceofSfilter} gives the performance of various indexes in a local computation node. The Bloom filter is tested using  breeze~\footnote{\url{https://github.com/scalanlp/breeze}}.
The sFilter(ad) represents the sFilter with adaptation of its structure given changes in the queries, and with the merging to reduce its size as introduced in Section~\ref{section:sfilterAdaptive}.
From Table~\ref{tab:performanceofSfilter}, observe that sFilter achieves one and two orders of magnitude speedup over the  Quadtree- and R-tree-based approaches in terms of spatial range search execution time, respectively. 
The sFilter(ad) improves the query processing time over the approach without optimization, but the sFilter(ad) has the overhead to merge branches of the index to control its size, and increases the false positive ratio. The Bloom filter does not support spatial range queries. 
Table~\ref{tab:performanceofSfilter} also gives the space usage overhead for various local indexes in each worker. The sFilter is 5-6 orders of magnitude less than the other types of indexes, e.g., the R-tree and the Quadtree. This is due to the bit encoding of the sFilter that eliminates the need for pointers. 
Moreover, the sFilter reduces the unnecessary network communication. We study the shuffle cost for redistributing the queries. The results are given in Figure~\ref{fig:effect_of_sfilter}. The sFilter reduces the shuffling cost for both spatial range and $k$NN join operations. The shuffle cost reduction depends on the data and query distribution. Thus, the more unbalanced  
the distribution of queries and data in the various computation nodes, the more shuffle cost is reduced. For example, for $k$NN join based on~\system, the shuffle cost is improved from 1114575 to 928933 when $k$ is 30, achieving 18\% reduction in network communication cost. The sFilter also reduce the network communication cost for Simba

\begin{table}[htbp]
\centering
\scalebox{0.80}{
\begin{tabular}{llllll} \hline
Dataset
& Index
&\vtop{\hbox{\strut Query}\hbox{\strut time(ms)}}
&\vtop{\hbox{\strut Index}\hbox{\strut build(s)}}
&\vtop{\hbox{\strut False}\hbox{\strut positive }}
&\vtop{\hbox{\strut Memory}\hbox{\strut usage(MB)}}
\\
\hline
\multirow{5}{*}{\centering Twitter}
&R-tree&19&17&/&112\\
&Q-tree&0.4&1.8&/&37\\
&sFilter&0.022&2&0.07&0.006\\
&sFilter(ad)&0.018&2.3&0.09&0.003\\
&Bloom filter&0.004&1.54&0.01&140\\
\hline
\multirow{5}{*}{\centering OSMP}
&R-tree&4&32&/&170\\
&Q-tree&0.5&1.2&/&63\\
&sFilter&0.008&2.4&0.06&0.008\\
&sFilter(ad)&0.006&6&0.10&0.006\\
&Bloom filter&0.002&2.7&0.01&180\\
\hline
\end{tabular}
}
\caption{Performance of the sFilter}
\label{tab:performanceofSfilter}
\end{table}

\subsection{Effect of the Number of Workers}
\begin{figure}
        \centering
        \begin{subfigure}[b]{0.19\textwidth}
                \includegraphics[width=\columnwidth]{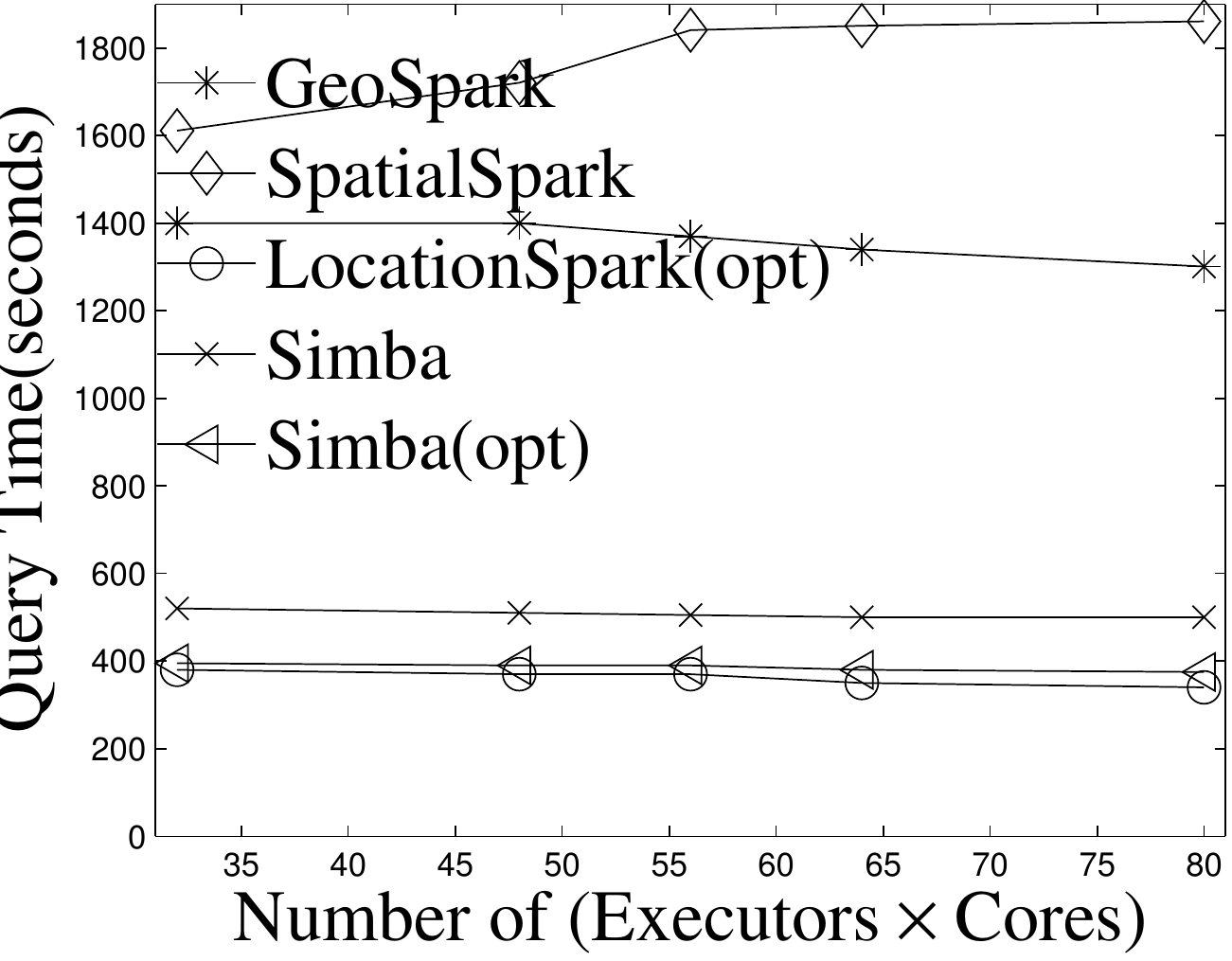}
                \caption{Spatial range join}
                \label{fig:knnjoin_datasizetwitter}
        \end{subfigure}%
        \begin{subfigure}[b]{0.19\textwidth}
               \includegraphics[width=\columnwidth]{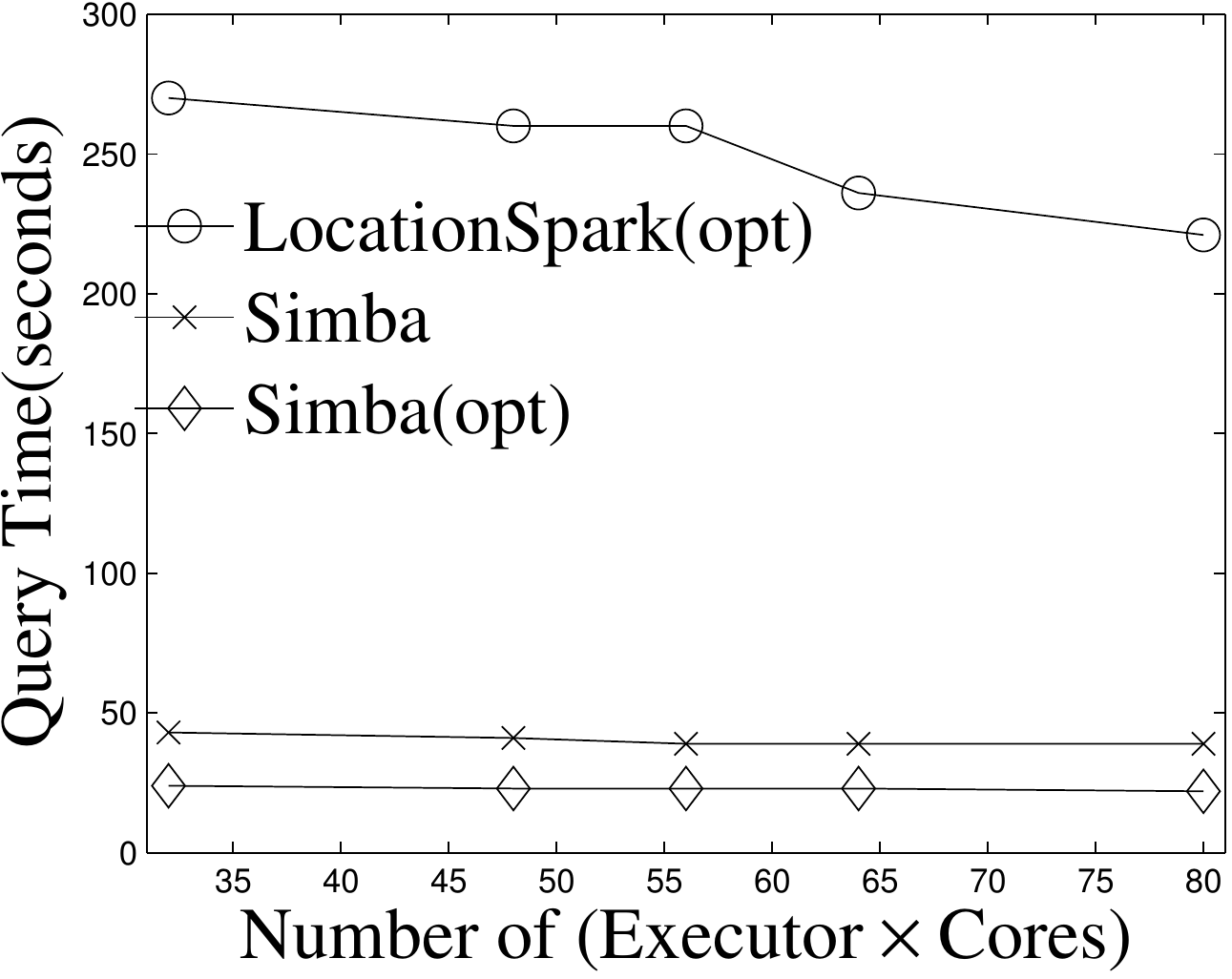}
                \caption{$k$NN join}
                \label{fig:knnjoin_twitter_numberofexecutors}
        \end{subfigure}
        \caption{Performance of spatial range join and $k$NN join when varying the number of executors}
        \label{fig:numberofexecutors}
\end{figure}

Spark's parallel computation ability depends on the number of executors and number of CPU cores assigned to each executor, that is, the number of executors times 
the number of CPU cores per executor. 
Therefore, to demonstrate the scalability of the proposed approach, we change the number of executors from 4 to 10, and fix the number of CPU cores assigned to each executor. We  study the runtime performance for spatial range join and $k$NN join operations using the Twitter and OSMP datasets on the Amazon EC2 cluster. Because the corresponding performance on the OSMP dataset gives similar trends as the
Twitter dataset, we only present the performance for the Twitter dataset in Figure~\ref{fig:numberofexecutors}, where the outer table size 
is fixed to 1 million around Chicago area, and the inner table size is 15 million. We observe that 
%WGA: Why is the size of the outer table larger than the size of the inner table? See above.
%MIngjie: this is typro. they placed in wrong direction. 
the performance of~\system and Simba(opt) for the spatial range join and the $k$NN join improves gradually with the increase in the number of executors. In contrast, GeoSpark and SpatialSpark do not scale well in comparison to~\system for spatial range join. The performance of Magellan for spatial join is not shown because it is based on Cartesian product and shows the worst performance. 

\section{Related Work}
\label{section:relatework}
\begin{sloppypar}
Spatial data management has been extensively studied for decades and several surveys provide good overviews. Gaede and G\"{u}nther~\cite{Gaede:Survey} provide a good summary of spatial data indexing. Sowell et al. give a survey and experimental study of iterative spatial-join in memory~\cite{SpatialjoinSurevy}. 
Recently, there has been considerable interest in supporting spatial data management over Hadoop MapReduce. 
%Afrati and Ullman~\cite{ilprints957} have proposed a framework that computes a multi-join query in a single computation round.
%Liu et al.~\cite{PGBJ} study how to use the Voronoi diagram data partition to speedup $k$NN join over MapReduce.
Hadoop-GIS~\cite{HadoopGIS} supports spatial queries in Hadoop by using a uniform grid index. SpatialHadoop~\cite{spatialhadoop} builds global and local spatial indexes, and modifies the HDFS record reader to read data more efficiently. 
MD-Hbase~\cite{MDHbase} extends HBase to support spatial data update and queries. Hadoop MapReduce is good at data processing for high throughput and fault-tolerance.
%Yet, Hadoop MapReduce has to write intermediate data into HDFS, and hence impedes the performance of applications that require pipelines of multiple MapReduce jobs.

%in-memory spatial database
Taking advantage of the very large memory pools available in modern machines, Spark and Spark-related systems (e.g., Graphx, Spark-SQL, and DStream)~\cite{Graphx,ZahariaPHDThesis} are developed to overcome the drawbacks of MapReduce in specific application domains. In order to process big spatial data more efficiently, it is natural to develop an efficient spatial data management systems based on Spark.  Several prototypes have been proposed to support spatial operations over Spark, e.g., GeoSpark~\cite{GeoSpark},   SpatialSpark~\cite{SpatialSpark},  Magellan~\cite{Magellan}, Simba~\cite{Simba}. However, some important factors impede the performance of these systems, mainly, query skew, lack of adaptivity, and excessive and unoptimized network and I/O communication overheads.
%are not addressed in these systems. 
For existing spatial join~\cite{SpatialjoinSurevy,DualTree} and $k$NN join approaches~\cite{Gorder,tkdeAllkNN,PGBJ}, we conduct experiments to study their performance in Section~\ref{section:locatlPlan}. The reader is referred to Section~\ref{section:locatlPlan}. For testing how the proposed techniques to improve Simba, we placed the spatial query scheduler ans sFilter into Simba standlone version. The Simba standlone version is based on Spark SQL Dataframe while removing the support of Spark SQL parser. The query scheduler and sFilter is placed into the physical plan of Simba (e.g., RDJSpark and RKJSpark).   

Kwon \textit{et al.}~\cite{SkewTune,SkewHandler} propose a skew handler to address the computation skew in a MapReduce platform. AQWA~\cite{AlyMHAOEQ15}  is a disk-based approach that handles spatial computation skew in MapReduce. In \system, we overcome the spatial query skew for spatial range join and $k$NN join operators, and provide an optimized query execution plan. These operators are not addressed in AQWA. The query planner in \system is different from relational query planners, i.e., join order and selection estimation. 
ARF~\cite{rangeFilter} supports one dimensional range query filter for data in disk. Calderoni \textit{et al.}~\cite{SBF} study spatial Bloom filter for private data. Yet, it does not support spatial range querying.
\end{sloppypar}

\section{Conclusions}
\label{section:conclusion}
We presented a query executor and optimizer to improve the query execution plan generated for spatial queries. 
%We conduct an extensive experimental study for local execution plan generation. 
We introduce a new spatial bitmap filter to reduce the redundant network communication cost.
Empirical studies on various real datasets  demonstrate the superiority of our approaches compared with existing systems. 

\section{Acknowledgement}
This work is supported by the National Science Foundation under Grant Number III- 1815796. 

\bibliographystyle{abbrv}
\bibliography{sigproc}

 \vspace{-18 mm}
\begin{IEEEbiography}[]
    {Mingjie Tang} is member of technical stuff at Hortonworks. He won his PhD at Purdue University. His research interests include
database system and big data infrastructure. He has an MS in computer science
from the University of Chinese Academy of Sciences, BS degree from Sichuan University, China.
\end{IEEEbiography}

 \vspace{-17 mm}
 
\begin{IEEEbiography}
    {Yongyang Yu} is a PhD student at Purdue University. His research focus is on big data management and matrix computation.
\end{IEEEbiography}
 
 \vspace{-17 mm}
 
\begin{IEEEbiography}
    {Walid G. Aref}
is a professor of computer science at Purdue. He is an associate editor of the ACM Transactions of Database Systems (ACM TODS) and has been an editor of the VLDB Journal. He is a senior member of the IEEE, and a member of the ACM. Professor Aref is an executive committee member and the past chair of the ACM Special Interest Group on Spatial Information (SIGSPATIAL).
\end{IEEEbiography}

 \vspace{-17 mm}
\begin{IEEEbiography}
    {Ahmed R.Mahmood} is a PhD student at Purdue University. His research focus is on big spatial-keyword data management.
\end{IEEEbiography}

 \vspace{-17 mm}
 
\begin{IEEEbiography}
    {Qutaibah~M.~Malluhi} joined Qatar University in September 2005. He is the Director of the KINDI Lab for Computing Research. He served as the head of Computer Science and Engineering Department at Qatar University between 2005-2012. 
\end{IEEEbiography}

  \vspace{-17 mm}
 
\begin{IEEEbiography}
    {Mourad Ouzzani}
is a Principal Scientist with the Qatar Computing Research
Institute (QCRI), Qatar Foundation. He is interested in research
and development related to data management and scientific data and
how they enable discovery and innovation in science and engineering.
Mourad received the Purdue University Seeds of
Success Award in 2009 and 2012. 
\end{IEEEbiography}

\begin{appendix}
\label{section:appendix}

\subsection{Proof of data repartition and query plan optimization problem}

\begin{sloppypar}
\begin{proof}
We prove it by reduction from the Bin-Packing problem~\cite{Garey:NP-hard}. Bin-Packing is defined as giving the finite set $U$ of items, the size of each item, says $s(u)$, is a positive real number, while $u \in U$, and a positive integer bin capacity $B$, and a positive Integer $M$, we want to find a partition of $U$ into disjoint set $U_1, U_2, ..., U_m$ such that the sum of the size of items in each $U_i$ is $B$ or less. Let $D_1', D_2', \ldots, D_k'$ be the solution to the reduced optimal data repartition problem. Next we show that based on $D_1', D_2', \ldots, D_k'$, we can obtain the solution to the original Bin-Packing problem in polynomial time.

To reduce this Bin-Packing problem to optimal data repartition problem, we first assume the skew and non-skew set is known. In practical, we can sort data partitions based on their approximate runtime $\E(D_i)$, and then find skewed partitions based on certain threshold. This takes polynomial time. Then the runtime over non-skewed partition is $\max_{j \in [ 1, \bar{N} ]} \{ \E(\bar{D_j}) \} = \Delta $, and is a constant value. Meanwhile, we disregard the minimization condition over $\bar{\rho (Q)}$. Therefore, from Equation~\ref{eq:opTcostfunction}, we can simplify this data repartition problem as
\begin{equation}\label{eq:minizecostfunction}
\minimize \{ \max \{ \max_{i \in [ 1, \hat{N} ]}  \{\widehat{\E(D_i^{s})} \}, \Delta  \} \}\}
\end{equation}
From this formula, we can find $\Delta$ is a constant value, thus, we have to minimize our cost function to be smaller than $\Delta$, that is,
\begin{equation}\label{eq:upperbound}
\minimize \{  \max_{i \in [ 1, \hat{N} ]}  \{\widehat{\E(D_i^{s})} \} \} \} \leq  \Delta
\end{equation}
From Equation~\ref{eq:oneskewcostfunction}, we can update Equation~\ref{eq:upperbound} to be
%MO: Add a new line! 
\begin{sloppypar}
\begin{equation} \label{eq:updatecostfunction}
\minimize \{  \max_{i \in [ 1, \hat{N} ]}  \{\beta(D_i) +  \max_{s \in [1, m']} \{\gamma(D_s)+ \E(D_s)\} \}+ \text{M}(Q_i) \} \} \} \leq  \Delta
\end{equation}
\end{sloppypar}
To simplify this function further, we disregard the optimization over $\beta(D_i), \gamma(D_s), text{M}(Q_i)$, we can simply our problem as following
\begin{sloppypar}
\begin{equation} \label{eq:updatecostfunction_1}
\minimize \{  \max_{i \in [ 1, \hat{N} ]}  \{ \max_{s \in [1, m']} \{\E(D_s)\} \}\}  \leq  \Delta
\end{equation}
\end{sloppypar}
Suppose the total number of new generated data partition is $k'$, this function is written as this way.
\begin{equation} \label{eq:updatecostfunction_2}
\minimize \{  \max_{s \in [ 1, k' ]}   \{\E(D_s)\} \}  \leq  \Delta
\end{equation}
Therefore, if $\E(D_s)  \leq  \Delta$, when $s \in [1, k']$, we can make sure our optimal function is at least less than its upper bound.

In addition, we assume that optimizer could split the skew data partitions into $m'$ partitions and $m'>M$, and each new computed data partition $D_i^{s}$ is associated with the cost $\widehat{\E(D_i^{s})}$. Now, optimizer can pack the computed partitions $D_i^{s}$ into a disjoint set, s.t., (1) the size of disjoint set is $M$, because $M$ is maximum number of data partitions. (2) the sum of cost in each bin would be smaller than $\Delta$ based on Equation~\ref{eq:updatecostfunction_2}. From this way, we can find this is same as the Bin-Packing problem. Thus, if $D_1', D_2', \ldots, D_k'$ is the solution of the optimal data repartition problem, then, we can use $D_1', D_2', \ldots, D_k'$ to compute the solution for the Bin-Packing problem. Since Bin-Packing is known to be NP-complete, and Bin-Packing is easier than optimal data repartition problem, we can deduce that optimal data repartition problem is also NP-complete.
\end{proof}
\end{sloppypar}

%\subsection{Simba with query scheduler and sFilter}

%Spatial distance join used the RDJ-spark. 
%  * Distance Join based on Two-Level R-Tree Structure
%  * (1) partitionted the query points and build a R-tree
%  * (2) duplicate the data points with the built R-tree
%  * (3) repatition the duplicated datapoints with the same data partition number as the query points
%  * (4) nest-loop local computation

\end{appendix}

\end{document}